\documentclass[a4paper]{article}

\usepackage[utf8]{inputenc}
\usepackage[english]{babel}
\usepackage{csquotes}
\usepackage{graphicx}
\usepackage{a4wide}
\usepackage{parskip}
\usepackage{amssymb, amsmath}
\usepackage[amsmath,hyperref,thmmarks]{ntheorem}

\usepackage{enumerate}
\usepackage{hyperref}
\usepackage{color,caption}
\usepackage{mathtools}
\usepackage{bm}
\usepackage{enumitem}

\usepackage{booktabs}
\usepackage{multirow}

\usepackage{tikz-cd}
\usepackage{tikz}
\usetikzlibrary{positioning}

\usepackage{pgfplots}
\pgfplotsset{compat=1.18}

\usepackage[ruled,vlined]{algorithm2e}
\SetKwInput{KwInit}{Initialization}
\SetAlgoNoLine
\DontPrintSemicolon

\usepackage{float}
\usepackage[section]{placeins}

\usepackage[
  backend=bibtex,
  style=numeric
]{biblatex}
\numberwithin{equation}{section}

\def\cF{{\mathcal F}}
\def\R{{\mathbb R}}

\newcommand{\Xb}{\mathbf{X}}
\newcommand{\Rb}{\mathbf{R}}

\newcommand{\balpha}{\bm{\alpha}}
\newcommand{\btheta}{\bm{\theta}}
\newcommand{\E}{\mathbb{E}}

\def \DCVaR{\mathrm{DCVaR}}
\def \CVaR{\mathrm{CVaR}}
\def \VaR{\mathrm{VaR}}
\def \APV{\mathrm{APV}}
\def \CF{\mathrm{CF}}
\def \BEL{\mathrm{BEL}}

\DeclareMathOperator{\Var}{Var}

\newtheorem{theorem}{Theorem}[section]
\newtheorem{proposition}[theorem]{Proposition}

\newenvironment{myproof}{\textit{Proof}.}{\hfill$\square$}

\newtheorem{remark}[theorem]{Remark}

\begin{document}

\date{\small \today}

\title{Multi periods mean-DCVaR optimization: a Recursive Neural Network resolution}

\author{Jérôme Lelong \thanks{Univ. Grenoble Alpes, CNRS, Grenoble INP, LJK, 38000 Grenoble, France. \texttt{jerome.lelong@univ-grenoble-alpes.fr}} \ \ Véronique Maume-Deschamps \thanks{Universite Lyon 1, Ecole Centrale de Lyon, INSA Lyon, Université Jean Monnet, CNRS, ICJ UMR5208,
69622 Villeurbanne, France. \texttt{veronique.maume-deschamps@univ-lyon1.fr}} \\  William Thevenot \thanks{Universite Lyon 1, Ecole Centrale de Lyon, INSA Lyon, Université Jean Monnet, CNRS, ICJ UMR5208,
        69622 Villeurbanne, France. and Risk Knowledge team at SCOR SE, Paris, France \texttt{thevenot.william@hotmail.com}} }

\maketitle


\begin{abstract}
  We study a discrete-time multi-period portfolio optimization problem under an explicit
  constraint on the Deviation Conditional Value-at-Risk (DCVaR), defined as the excess of
  Conditional Value-at-Risk over expected terminal wealth. The objective is to maximize
  expected return subject to a global tail-risk constraint, leading to a time-inconsistent
  precommitment problem. We propose a recurrent neural-network-based approach to approximate
  the optimal precommitment policy, which accommodates path-dependent risk constraints and
  high-dimensional state dynamics without relying on dynamic programming. The explicit
  constraint formulation allows for exact penalty methods and provides a transparent notion
  of feasibility. The methodology is validated in a classical complete-market financial model
  and extended to a multi-period portfolio allocation problem in (re)insurance, capturing
  the long-term risk dynamics of insurance liabilities.
\end{abstract}


\section{Introduction}

The mean-CVaR portfolio optimization framework, which involves maximizing expected return under Conditional Value-at-Risk constraints, became widely studied following the seminal contributions of~\textcite{rockafellar2000optimization} and~\textcite{krokhmal2002portfolio}.  
A key limitation of this approach lies in the time-inconsistency of the CVaR measure, highlighted by~\textcite{artzner2007coherent} and rigorously established in~\textcite{shapiro2009time}, which prevents the direct application of dynamic programming and standard stochastic control techniques.

An alternative perspective is provided by the deviation-measure framework developed by Rockafellar, Uryasev, and co-authors~\cite{Rockafellar2002Deviation,rockafellar2006optimality,rockafellar2006generalized,rockafellar2013fundamental}, where dispersion is quantified around the expectation and linked to coherent risk measures via $\rho(X)=\E[X]-D(X)$. 
This leads to the mean--deviation formulation, a convex extension of the classical mean--variance paradigm encompassing variance, semi-deviation, CVaR, and DCVaR. 
In contrast to mean--risk formulations, which may fail to be well posed when expected returns are unbounded, the mean--deviation problem remains coercive under standard assumptions and therefore always admits an optimal solution. 
This viewpoint naturally motivates the use of DCVaR as a central risk functional.

To address mean--(D)CVaR optimization problems, various approaches have been proposed.  
In the discrete-time setting,~\textcite{strub2019discrete} reformulated the mean-CVaR problem as a collection of expected-utility maximization problems with piecewise linear utility functions, thereby recovering time-consistency and maintaining analytical tractability.  
In the continuous-time case,~\textcite{gao2017dynamic} employed a martingale approach within a complete market, converting the dynamic optimization into a static one under the risk-neutral measure.  
Along similar lines,~\textcite{lelong2025martingale} analyzed the mean-DCVaR problem and obtained explicit solutions under comparable assumptions, such as bounded terminal wealth.

Another line of research, initiated by~\textcite{miller2017optimal}, addresses the mean-CVaR control problem in continuous time through a bilevel optimization formulation.  
Their framework avoids the need for a complete market and naturally incorporates time-inconsistency via a nested optimization structure, albeit at the cost of solving a sequence of value-function problems.

Despite these advances, existing approaches typically rely on problem-specific reformulations or dynamic programming techniques that become computationally prohibitive in high-dimensional or constrained settings.

In this work, we investigate the discrete-time mean-DCVaR problem with an \emph{explicit} risk constraint and propose a global optimization approach based on neural network parameterizations. The control policy is represented by a recurrent neural network, allowing us to directly optimize the multi-period objective under a terminal DCVaR constraint without relying on backward recursion. The proposed methodology is applied both to a classical financial setting with a complete market and extended to a multi-period portfolio allocation problem in (re)insurance, designed to capture the long-term risk dynamics of insurance liabilities. In this context, the framework provides a basis for building strategic guidance tools to structure risk–return trade-offs under uncertainty, rather than to produce precise forecasts. Our framework builds on recent advances in neural-network-based portfolio optimization~\textcite{li2019data,van2024global,ni2025optimal}, and extends them to the mean--DCVaR setting.

The choice of an explicit constraint formulation, rather than a penalized risk-aversion approach, is motivated by both methodological and practical considerations.
From a methodological standpoint, it preserves the target level \(K\) within the optimization problem and allows us to leverage Clarke’s exact penalty principle~\cite{clarke1983optimization}, ensuring equivalence between constrained and penalized formulations beyond a finite threshold.
From a practical perspective, explicit risk constraints arise naturally in insurance applications; in particular, solvency regulations such as the Solvency Capital Requirement (SCR) impose capital-at-risk limits that are structurally analogous to the DCVaR constraint considered here.

Finally, we emphasize that the learned policy is a \emph{precommitment} solution: the constraint is enforced at the initial date, and the resulting strategy may not be time-consistent. This viewpoint aligns with practical implementations, where a global risk threshold is fixed ex ante rather than enforced dynamically.

The paper is organized as follows. Section~2 introduces the mean--CVaR and mean--DCVaR formulations and discusses their relationship. Section~3 presents the multi-period problem. Section~4 describes the proposed neural network-based solution approach. Section~5 reports numerical experiments under a deterministic multidimensional Black--Scholes model. Section~6 extends the framework to (re)insurance portfolio optimization. An appendix provides the details of the simulation framework used for the life insurance portfolio in Section~6.

\section{Mean-DCVaR}

In this section, we introduce the mean-DCVaR formulation in a general setting, we discuss its relation to the mean-CVaR problem, and present some techniques for simplifying the search for optimal solutions.

\subsection{VaR, CVaR and DCVaR definitions}

For a real-valued random variable $X$ and a fixed confidence level $\kappa \in (0,1)$, we define the Value-at-Risk (VaR), the Conditional Value-at-Risk (CVaR), and the Deviation Conditional Value-at-Risk (DCVaR) as follows:
\begin{equation*}
    \begin{aligned}
         & \VaR_\kappa(X) = \inf \left\{ M \in \mathbb{R} : \mathbb{P}(X \le M) \ge \kappa \right\}, \\
         & \CVaR_\kappa(X) = \frac{1}{1 - \kappa} \int_\kappa^1 \VaR_{\kappa'}(X) \, d\kappa', \\
         & \DCVaR_\kappa(X) = \CVaR_\kappa\big(X - \E[X]\big) = \CVaR_\kappa(X)-\E[X].
    \end{aligned}
\end{equation*}

The CVaR is commonly used as a proxy for the VaR in optimization problems, as it satisfies the properties of a coherent risk measure in the sense of~\textcite{artzner2007coherent}.

Deviation measures were formally defined as coherent counterparts of risk measures by Rockafellar, Uryasev, and Zabarankin~\cite{Rockafellar2002Deviation}.
 
\subsection{Mean-CVaR and Mean-DCVaR formulations}

Let \(U:\mathbb{R}\to\mathbb{R}\) be measurable and let \(\{Y_{\boldsymbol{\alpha}};\,\boldsymbol{\alpha}\in\mathcal{A}\}\) be a family of real-valued random variables such that
\[
\mathbb{E}\bigl[|U(Y_{\boldsymbol{\alpha}})|\bigr] < \infty, 
\qquad \forall\,\boldsymbol{\alpha}\in\mathcal{A}.
\]
For a fixed confidence level \(\kappa \in (0,1)\), we consider the following constrained optimization problems:
\begin{equation}
\tag{$P_{\text{CVaR}}(K)$}\label{eq:PCVaR}
\inf_{\boldsymbol{\alpha}\in\mathcal{A}} \; -\E[U(Y_{\boldsymbol{\alpha}})]
\quad \text{subject to} \quad 
\CVaR_\kappa\bigl(-U(Y_{\boldsymbol{\alpha}})\bigr) \le K,
\end{equation}
\begin{equation}
\tag{$P_{\text{DCVaR}}(\tilde K)$}\label{eq:PDCVaR}
\inf_{\boldsymbol{\alpha}\in\mathcal{A}} \; -\E[U(Y_{\boldsymbol{\alpha}})]
\quad \text{subject to} \quad 
\DCVaR_\kappa\bigl(-U(Y_{\boldsymbol{\alpha}})\bigr) \le \tilde K.
\end{equation}

Problem~\eqref{eq:PCVaR} corresponds to the classical mean--CVaR formulation~\textcite{rockafellar2000optimization}, while~\eqref{eq:PDCVaR} defines its centered counterpart in terms of deviation. 
The latter removes the location component of the loss distribution and focuses on tail dispersion around the mean, a property that plays a key role in the well-posedness of the optimization problem. 

The relation between these two formulations is clarified by the following result, adapted from Theorem~4 in~\textcite{Rockafellar2002Deviation}.\\

\begin{theorem}[Inclusion of CVaR solutions in DCVaR solutions]\label{th:cvar_inclusion}
Assume that: $U$ is increasing, concave, and sublinear; $\mathcal{A}$ is convex; the mapping $\balpha \mapsto -U(Y_{\balpha})$ is convex.

Let $\balpha^\star$ be an optimal solution of problem~\eqref{eq:PCVaR} for a given threshold $K$, 
such that the CVaR constraint is \emph{active}, that is, $\CVaR_\kappa(-U(Y_{\balpha^\star})) = K$.
Define the corresponding calibrated threshold for the DCVaR problem as
\[
\tilde K \;=\; K + \E\!\big[U(Y_{\balpha^\star})\big].
\]
Then, the problems~\eqref{eq:PCVaR} and~\eqref{eq:PDCVaR} share the same set of optimal solutions 
and the same optimal objective value.
\end{theorem}

To establish the result, we rely on the following theorem from~\textcite{krokhmal2002portfolio}, 
which recalls the conditions under which the three optimization formulations \((P1)\)–\((P3)\) below are equivalent.

\begin{theorem} \cite{krokhmal2002portfolio}\label{th:three_formulation}
Let $\mathcal A$ be a set, and let
$R:\mathcal A \to \mathbb R$ and $\phi:\mathcal A \to \mathbb R$
denote respectively an expected return functional and a risk measure.
Consider the following three optimization problems:
\[
\begin{aligned}
\text{(P1)}\;&\quad \inf_{\balpha \in \mathcal{A}}\; -R(\balpha)+ \lambda \phi(\balpha),
&& \lambda \ge 0,\\[0.3em]
\text{(P2)}\;&\quad \inf_{\balpha \in \mathcal{A}}\; \phi(\balpha),
&& R(\balpha) \ge \mu,\\[0.3em]
\text{(P3)}\;&\quad \inf_{\balpha \in \mathcal{A}}\; -R(\balpha),
&& \phi(\balpha) \le K.
\end{aligned}
\]
Assume that the constraints $R(\balpha) \ge \mu$ and $\phi(\balpha) \le K$
admit strictly feasible points. If $\phi$ is convex, $R$ is concave, and $\mathcal A$ is convex,
then varying the parameters $\lambda$, $\mu$, and $K$ parametrizes the
same efficient frontier, and the three formulations (P1)--(P3) are
equivalent in the sense that they generate identical sets of efficient
solutions.\\
\end{theorem}

\begin{myproof}[Proof of Theorem~\ref{th:cvar_inclusion}]
Let $\balpha^\star$ be an optimal solution to $P_{\text{CVaR}}(K)$ with an active CVaR constraint, that is,
\[
\CVaR_\kappa(-U(Y_{\balpha^\star})) = K.
\]

\medskip
By Theorem~\ref{th:three_formulation}, there exists $\lambda_{\mathrm C}\ge 0$ such that $\balpha^\star$ minimizes the penalized functional over $\mathcal{A}$:
\[
F_{\mathrm C}(\balpha) := -\E[U(Y_{\balpha})] + \lambda_{\mathrm C}\, \CVaR_\kappa(-U(Y_{\balpha})),
\qquad \text{(CVaR penalization).}
\]

\medskip
Set $\lambda_{\mathrm D} = \lambda_{\mathrm C}/(1+\lambda_{\mathrm C}) \in [0,1)$, so that $\lambda_{\mathrm C} = \lambda_{\mathrm D}/(1-\lambda_{\mathrm D})$.

Then, we have
\begin{align*}
(1-\lambda_{\mathrm D}) F_{\mathrm C}(\balpha) 
  &= -(1-\lambda_{\mathrm D})\, \E[U(Y_{\balpha})] 
     + \lambda_{\mathrm D}\, \CVaR_\kappa(-U(Y_{\balpha})) \\
  &= -\E[U(Y_{\balpha})] 
     + \lambda_{\mathrm D}\big(\CVaR_\kappa(-U(Y_{\balpha})) + \E[U(Y_{\balpha})]\big) \\
  &= -\E[U(Y_{\balpha})] 
     + \lambda_{\mathrm D}\, \DCVaR_\kappa(-U(Y_{\balpha})).
\end{align*}

Since $(1-\lambda_{\mathrm D}) > 0$, minimizing 
\[
F_{\mathrm D}(\balpha) := -\E[U(Y_{\balpha})] + \lambda_{\mathrm D}\, \DCVaR_\kappa(-U(Y_{\balpha}))
\]
is equivalent to minimizing $F_{\mathrm C}(\balpha)$. 

\medskip
Therefore, there exists a threshold $\tilde{K}$ such that $\balpha^\star$ minimizes $P_{\text{DCVaR}}(\tilde K)$ by Theorem~\ref{th:three_formulation}. 
Moreover,
\[
\tilde K = \DCVaR_\kappa(-U(Y_{\balpha^\star})) 
         = K + \E[U(Y_{\balpha^\star})].
\]
This completes the proof.

\end{myproof}

\begin{remark}[Mean--CVaR vs.\ Mean--DCVaR]
It should be noted that the scalarized mean--DCVaR and
mean--CVaR problems are ordered for all values of the penalization
parameter. When $\lambda_{\mathrm D} \in [0,1)$, the two scalarized
formulations are equivalent and therefore generate the same efficient
frontier. Outside this range, the problems differ in nature and no
longer describe the same mean--risk trade-off.

For each risk specification separately, Theorem~\ref{th:three_formulation}
ensures the equivalence between the scalarized and constrained
formulations (P1)--(P3) under the stated convexity and feasibility
assumptions.

Accordingly, the deviation-based formulation contains the mean--CVaR
problem as a particular case and provides a genuine extension of the
classical mean--CVaR framework. This structural relation is discussed
in Rockafellar, Uryasev, and Zabarankin~\cite{rockafellar2006generalized}.
\end{remark}

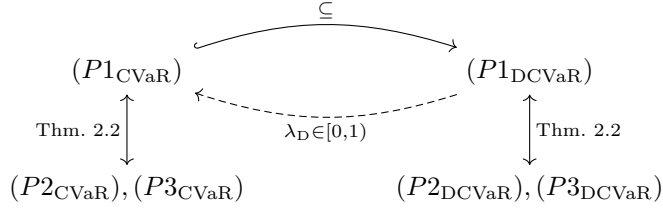
\begin{figure}[H]
\centering
\begin{tikzcd}[column sep=huge, row sep=large]
(P1_{\mathrm{CVaR}})
\arrow[r, hook, bend left=18, "\subseteq"]
\arrow[d, leftrightarrow, "\text{Thm.~\ref{th:three_formulation}}"']
&
(P1_{\mathrm{DCVaR}})
\arrow[l, dashed, bend left=18, "{\lambda_{\mathrm D}\in[0,1)}"]
\arrow[d, leftrightarrow, "\text{Thm.~\ref{th:three_formulation}}"]
\\
(P2_{\mathrm{CVaR}}),(P3_{\mathrm{CVaR}})
&
(P2_{\mathrm{DCVaR}}),(P3_{\mathrm{DCVaR}})
\end{tikzcd}
\caption{Relations between the mean--CVaR and mean--DCVaR formulations.}
\label{fig:cvar_dcvar_relations}
\end{figure}

\subsection{Auxiliary variable formulation}

For any \(\balpha \in \mathcal{A}\) and \(\eta \in \mathbb{R}\), define
\[
g(\balpha,\eta)
:= \eta + \frac{1}{1-\kappa}\,\E\!\left[\big(-U(Y_{\balpha}) - \eta\big)_+\right] + \E[U(Y_{\balpha})].
\]

Assume that for any \(\balpha \in \mathcal{A}\), the function \(\eta \mapsto g(\balpha,\eta)\) is convex and continuously differentiable. Then, for any \(\balpha\),
\[
D(\balpha) = \inf_{\eta \in \mathbb{R}} g(\balpha,\eta),
\qquad 
A(\balpha) := \arg\min_{\eta \in \mathbb{R}} g(\balpha,\eta).
\]
We then define
\[
\phi(\balpha) := \min\{\eta \in A(\balpha)\},
\qquad 
D(\balpha) = g(\balpha,\phi(\balpha)).
\]

The following result provides an equivalent lifted formulation of the DCVaR-constrained problem.

\begin{proposition}\label{prop:aux_pb}
Assume that \(\E[|U(Y_{\balpha})|] < \infty\) for all \(\balpha \in \mathcal{A}\).
Then, for any \(K \in \mathbb{R}\), the problems
\[
\inf_{\balpha \in \mathcal{A}} \; -\E[U(Y_{\balpha})]
\quad \text{subject to} \quad D(\balpha) \le K,
\]
and
\[
\inf_{(\balpha,\eta) \in \mathcal{A} \times \mathbb{R}} \; -\E[U(Y_{\balpha})]
\quad \text{subject to} \quad g(\balpha,\eta) \le K
\]
are equivalent, in the sense that they share the same optimal value and that any optimal \(\balpha^\star\) can be completed into an optimal pair \((\balpha^\star,\eta^\star)\) with \(\eta^\star \in A(\balpha^\star)\).
\end{proposition}

This result is classical for the mean--CVaR formulation~\textcite{rockafellar2000optimization}.  
The extension to the mean--DCVaR case follows directly from the identity
\(D(\balpha) = \CVaR_\kappa(-U(Y_{\balpha})) + \E[U(Y_{\balpha})]\),
using the same arguments.

\section{Multi-period problem}

We first present a standard financial market framework used to construct and validate the neural network approach. In Section~\ref{sec:reinsurance}, we extend the analysis to an insurance setting with dependent returns, non self-financing strategies, and short horizons ($N \le 10$).

We consider a discrete-time model on a filtered probability space 
$(\Omega,\mathbb F = (\cF_n)_{0\le n\le N},\mathbb P)$.
An investor with initial wealth \(V_0\) allocates wealth among one risk-free asset and \(d\) risky assets, with rebalancing dates \(t_0,\ldots,t_N\).

Let \((\alpha_n^0,\balpha_n)_{n=1}^N\) denote the holdings in the risk-free and risky assets, where the control process \(\balpha=(\balpha_n)_{n=1}^N\) is \(\mathbb F\)-predictable, so that \(\balpha_{n+1}\) is chosen at time \(t_n\).

Denote by \(\Rb_n=(R_n^1,\ldots,R_n^d)\) the vector of risky gross returns and by \(R_n^0\) the risk-free return, with
\[
R_n^i = \frac{X_n^i}{X_{n-1}^i}, \qquad R_n^i > 0 \ \text{a.s.}
\]
No specific distributional assumption is imposed on \((\Rb_n)_{1\le n\le N}\).

We assume that the returns are conditionally independent given the previous state \(S_{n-1}\), i.e.
\begin{equation}\label{conditional_indep_previous}
\Rb_n \;\perp\!\!\!\perp\; \{ \Rb_k : k < n \} \;\big|\; X_{n-1}.
\end{equation}

\medskip
Let \((\mathcal A_n)_{1\le n\le N}\) be a family of admissible sets, where each \(\mathcal A_n\) is an \(\mathcal F_{n-1}\)-measurable random set. 
We consider predictable controls \(\balpha=(\balpha_n)_{n=1}^N\) such that \(\balpha_n \in \mathcal A_n\).

Under the self-financing condition, the portfolio value evolves as
\begin{equation}\label{eq:self-financing}
V_{n+1}
=
\balpha_{n+1}^\top \Xb_{n+1}
+
\bigl(V_n-\balpha_{n+1}^\top \Xb_n\bigr) R_{n+1}^0,
\qquad n=0,\ldots,N-1.
\end{equation}

We consider the multi-period mean--DCVaR problem with a terminal constraint. 
Using Proposition~\ref{prop:aux_pb}, it can be equivalently written as
\begin{equation}\label{eq:PDCVaR_N_eq}
\begin{aligned}
\inf_{(\balpha_n)_{n=1}^N,\, \eta} \quad 
    & -\,\E[U(V_N)] \\
\text{s.t.} \quad 
    & g(\balpha_1,\ldots,\balpha_N,\eta) \le K, \\
    & \balpha_{n+1} \in \mathcal A_{n+1}, \quad
      \balpha_{n+1} \text{ is } \mathcal F_n\text{-measurable}, \\
    & V_{n+1}
      =
      \balpha_{n+1}^\top \Xb_{n+1}
      +
      \bigl(V_n - \balpha_{n+1}^\top \Xb_n\bigr) R_{n+1}^0,
      \qquad n=0,\ldots,N-1,
\end{aligned}
\end{equation}
where
\[
g(\balpha_1,\ldots,\balpha_N,\eta)
:= \eta + \frac{1}{1-\kappa}\,\E\!\left[\big(-U(V_N)-\eta\big)_+\right] + \E[U(V_N)].
\]

Under the standing assumptions that \(U\) is concave and nondecreasing, the admissible sets \(\mathcal A_n\) are convex, and the DCVaR functional is convex, the terminal wealth \(V_N\) is affine in the controls and the function \(g(\balpha_1,\ldots,\balpha_N,\eta)\) is convex in \((\balpha_1,\ldots,\balpha_N,\eta)\). 
Problem~\eqref{eq:PDCVaR_N_eq} is therefore convex. 
However, once the control is parameterized by a neural network \(A_\theta\), convexity in \(\balpha\) is lost, while convexity in \(\eta\) is preserved, leading to a nonconvex optimization problem in~\(\theta\).

\paragraph{Relation with Strub et al.~(2019).}

Strub et al.~\cite{strub2019discrete} show that, in a frictionless discrete-time setting, mean--CVaR problems can be reformulated as expected-utility maximization with a piecewise-linear utility, restoring time consistency and enabling dynamic programming.

In our setting, Theorem~\ref{th:cvar_inclusion} implies that, whenever the associated multiplier satisfies \(\lambda_D \in [0,1)\), the mean--DCVaR constrained problem admits an equivalent Lagrangian formulation of mean--CVaR type, with parameter
\[
\lambda_C = \frac{\lambda_D}{1-\lambda_D} > 0.
\]
The condition \(\lambda_D < 1\) ensures a strictly positive weight on the expectation term; for \(\lambda_D \ge 1\), this correspondence breaks down.

Thus, in the interior regime \(\lambda_D^\star \in [0,1)\), the results of Strub et al.~(2019) apply directly, the only difference being that the efficient frontier is parametrized here by a DCVaR constraint rather than by a mean constraint, making the expectation term endogenous to the deviation-based formulation.

\section{A global optimization approach with Neural Network}

We propose a global optimization framework based on neural network parameterizations to solve the multi-period mean--DCVaR problem. 
The control policy is represented by a recurrent neural network, allowing us to directly optimize the terminal objective under a DCVaR constraint without relying on dynamic programming.

The framework accommodates general nonlinear dynamics, high-dimensional state variables, and portfolio constraints (e.g., budget, leverage, or short-selling constraints), which can be enforced directly at the network output level. 
In practice, we employ a residual architecture to improve the stability and robustness of the learned allocation strategies.

\subsection{Neural network parameterization for the mean--DCVaR problem}

We aim to solve~\eqref{eq:PDCVaR_N_eq} by approximating the optimal control sequence \((\balpha_n)_{n=1}^N\) and the auxiliary variable \(\eta\). The controls are assumed to be \(\mathbb F\)-predictable.

Under our standing assumptions, the state process \((V_n,\Xb_n)_{n=0}^N\) is Markovian. Hence, without loss of optimality, the control can be chosen in feedback form
\[
\balpha_{n+1} = \balpha_{n+1}(V_n,\Xb_n).
\]

We parameterize the control in terms of portfolio proportions. 
For each time step, we introduce a neural network
\[
B_{n,\theta_n}:\mathbb{R}\times\mathbb{R}^d \to \mathbb{R}^{d+1},
\qquad 
\mathbf{1}^\top B_{n,\theta_n}(V_{n-1},\Xb_{n-1}) = 1,
\]
and define the corresponding holdings by
\[
\alpha_n^i
=
B_{n,\theta_n}^i(V_{n-1},\Xb_{n-1})\,\frac{V_{n-1}}{X_{n-1}^i},
\qquad i=0,\ldots,d.
\]
The parameters \((\theta_n)_{n=1}^N\) are trained jointly.

Portfolio constraints are enforced at the network output, e.g.
\[
B_{n,\theta_n}(V_{n-1},\Xb_{n-1}) \in \mathcal C(V_n),
\]
for some convex set \(\mathcal C(V_n)\) (e.g. simplex, leverage or short-selling constraints).

The resulting optimization problem reads
\begin{equation}
\label{eq:NN-DCVaR}
\begin{aligned}
\inf_{(\theta_n)_{n=1}^N,\;\eta\in\R} \quad
    & -\,\E[U(V_N)] \\
\text{s.t.} \quad
    & g(\balpha_1,\ldots,\balpha_N,\eta) \le K, \\
    & V_{n+1}
    =
    V_n \!\left(
      \sum_{i=1}^d B_{n+1,\theta_{n+1}}^i(V_n,\Xb_n)\, R_{n+1}^i
      + B_{n+1,\theta_{n+1}}^0(V_n,\Xb_n)\, R_{n+1}^0
    \right),\\
    & B_{n+1,\theta_{n+1}}(V_n,\Xb_n) \in \mathcal C(V_n),
    \qquad n=0,\ldots,N-1.
\end{aligned}
\end{equation}

This parametrization reduces the effective complexity of the learning problem while preserving the self-financing structure of the portfolio.

\subsection{Recurrent neural architecture with ResNet connections} \label{subsec:RNN}

To reduce the number of parameters, we use a recurrent architecture with shared weights across time. 
Instead of training $N$ distinct networks, we introduce a single neural network
\[
B_{\btheta} : (V_n,\Xb_n,n) \longmapsto \mathbb{R}^{d+1},
\]
where \(\btheta\) denotes the shared parameters.~\eqref{eq:NN-DCVaR} can then be rewritten as
\begin{equation}
\label{eq:NNR}
\begin{aligned}
\inf_{\btheta,\eta} \quad 
    & -\,\E[U(V_N)] \\
\text{s.t.} \quad 
    & \eta + (1-\kappa)^{-1} 
      \E\!\left[(-U(V_N)-\eta)_+\right]
      + \E[U(V_N)]
      \le K, \\
    & V_{n+1}
      =
      V_n \!\left(
        \sum_{i=1}^d B_{\btheta}^i(V_n,\Xb_n,n)\, R_{n+1}^i
        + B_{\btheta}^0(V_n,\Xb_n,n)\, R_{n+1}^0
      \right), \\
    & B_{\btheta}(V_n,\Xb_n,n) \in \mathcal C(V_n),
    \qquad n=0,\ldots,N-1.
\end{aligned}
\end{equation}

The network takes as input \((V_n,\Xb_n,n)\) and outputs portfolio weights in \(\mathbb{R}^{d+1}\). 
A projection layer (e.g., softmax) enforces admissibility by ensuring non-negativity and unit sum.

The control policy is parameterized by a recurrent neural network (RNN), where at each time step the update is given by a feedforward neural network \(B_{\btheta}\).

To improve stability for large horizons, we incorporate residual (ResNet-style) connections within \(B_{\btheta}\), so that each layer learns a correction of the form
\[
h_{k+1} = h_k + \varphi(W_k h_k + b_k).
\]
These skip connections mitigate vanishing-gradient issues and lead to more stable training, while preserving the overall recurrent structure of the policy.
\begin{figure}[h]
\centering
\begin{tikzpicture}[
  node distance=13mm,
  every node/.style={font=\small},
  box/.style={draw, rounded corners, inner sep=5pt, align=center},
  arrow/.style={->, >=stealth, line width=0.4pt}
]

\node[box] (state) {$\text{state at } t_n$\\$(V_n,\Xb_n,n)$};

\node[box, right=of state] (net)
{shared policy\\$B_{\btheta}(V_n,\Xb_n,n)$};

\node[box, right=of net] (proj)
{projection / constraints\\$\Pi_{\mathcal C(V_n)}$};

\node[box, right=of proj] (w)
{portfolio weights\\$w_{n+1}\in\mathcal C(V_n)$};

\draw[arrow] (state) -- (net);
\draw[arrow] (net) -- (proj);
\draw[arrow] (proj) -- (w);

\node[box, below=of proj] (dyn)
{wealth update\\
$V_{n+1}
=
V_n\!\left(\sum_{i=1}^d w_{n+1}^i R_{n+1}^i
+ w_{n+1}^0 R_{n+1}^0\right)$};

\draw[arrow] (w.south) -- (dyn.north);

\draw[arrow]
  (dyn.west) to[out=180,in=-90,looseness=0.9]
  node[midway,left] {$n\leftarrow n+1$}
  (state.south);

\node[box, below=of dyn] (terminal)
{\begin{tabular}{c}
terminal evaluation\\
$-\E[U(V_N)]$\\
$g(\btheta,\eta)\le K$
\end{tabular}};

\draw[arrow] (dyn.south) -- node[midway,right] {$V_N$} (terminal.north);

\end{tikzpicture}
\caption{Computational pipeline for the shared policy optimization in \eqref{eq:NNR}.}
\label{fig:pipeline_nnr}
\end{figure}
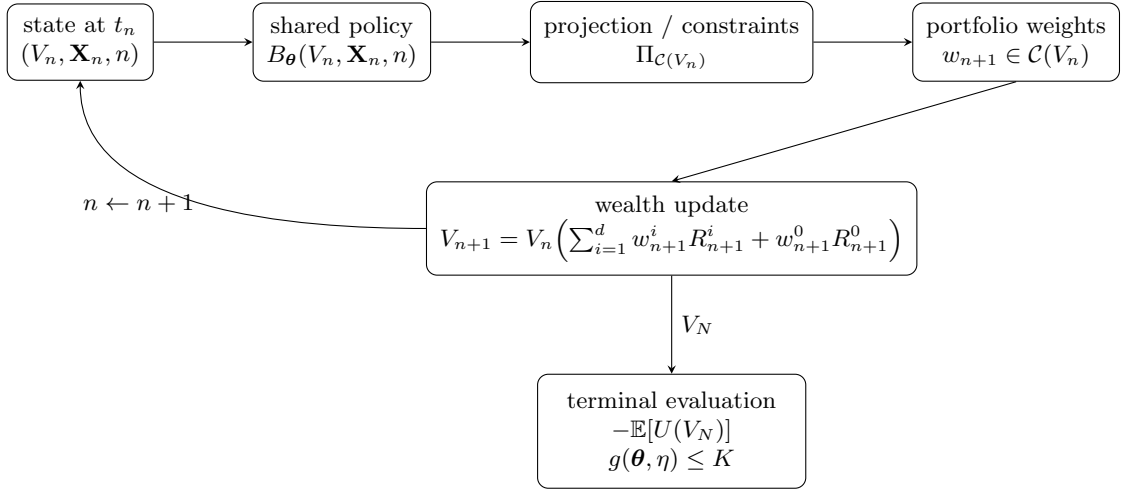

\paragraph{Constraints management.}

Portfolio constraints are enforced directly at the level of the trading rule by parameterizing the holdings as
\[
\alpha_{n+1}^i = w_{n+1}^i\,\frac{V_n}{X_n^i},
\]
where \(w_{n+1}=B_{\theta}(V_n,\Xb_n,n)\) denotes the portfolio weights. This parametrization preserves the self-financing condition by construction. 

Different constraint sets can be incorporated through the network output layer. 
Long-only and fully invested portfolios are obtained via a softmax mapping, ensuring \(w_{n+1}\in\Delta^d\). 
More general constraints, such as box constraints \(\ell \le w_{n+1} \le u\), are handled through a differentiable projection onto the feasible set 
\(\mathcal{C}=\{w:\ \mathbf{1}^\top w=1,\ \ell \le w \le u\}\). 
Unconstrained self-financing strategies can also be implemented by directly parameterizing portfolio values and enforcing the budget constraint through a residual allocation.

When short-selling is allowed, the portfolio value may become negative. In this case, we stop the trajectory at the first hitting time of zero (ruin) and use the corresponding value as terminal wealth.

\subsection{Training method}\label{sec:training_method}

We solve the constrained mean--DCVaR problem using an exact penalty approach. 
Let
\[
f(\btheta,\eta) := -\,\E[U(V_N^{\btheta})],
\qquad
g(\btheta,\eta)
:=
\eta + \frac{1}{1-\kappa}\E\!\left[(-U(V_N^{\btheta})-\eta)_+\right]
+ \E[U(V_N^{\btheta})],
\]
so that the constraint reads \(g(\btheta,\eta)\le K\).

We minimize the penalized objective
\begin{equation}
\label{eq:penalty-objective}
\mathcal{L}(\btheta,\eta)
=
f(\btheta,\eta)
+
\lambda_1\big(g(\btheta,\eta)-K\big)_+
+
\lambda_2\big((g(\btheta,\eta)-K)_+\big)^2,
\end{equation}
where \(\lambda_1,\lambda_2\ge0\). 
The linear term ensures exactness of the penalty in the sense of Clarke’s principle see~\textcite{clarke1983optimization}, while the quadratic term improves numerical stability. 
This formulation preserves the target level \(K\) explicitly, in contrast with classical risk-aversion penalizations~\textcite{rockafellar2000optimization,krokhmal2002portfolio}.

The penalty parameters are updated adaptively using projected gradient steps based on the constraint violation \(v_t := g(\btheta_t,\eta_t)-K\), while \((\btheta,\eta)\) are optimized via stochastic gradient methods (e.g.\ Adam) using Monte Carlo estimates.

In practice, gradient clipping and moving averages of the constraint violation are used to stabilize training near the constraint boundary. 
Optional calibration of the target level \(K\) can be performed to reduce residual constraint bias.

\begin{algorithm}[H]
\caption{Global training of the shared policy \(B_{\theta}\) for the mean--DCVaR problem}
\label{alg:nn_dcv}
\KwIn{Horizon \(N\), dimension \(d\), initial wealth \(V_0\), confidence level \(\kappa\), target \(K\), utility \(U\), Monte Carlo batch size \(M\), optimizer with stepsize \(\gamma\), EMA factor \(\beta\), dual stepsizes \((\rho_1,\rho_2)\), constraint set $\mathcal{C}$ with differentiable projection $\Pi_{\mathcal{C}}$.}
\KwInit{Initialize \(\theta\), \(\eta\), \(\lambda_1,\lambda_2 \ge 0\), and \(\bar v \leftarrow 0\).}

\For{iterations \(t=0,1,2,\dots\)}{
  \For{\(m=1,\dots,M\)}{
    \(V_0^{(m)} \leftarrow V_0\)\;
    \For{\(n=0,\dots,N-1\)}{
      \(w_{n+1}^{(m)} \leftarrow \Pi_{\mathcal C}\!\bigl(B_\theta(V_n^{(m)},\Xb_n^{(m)},n)\bigr)\)\;
      \(V_{n+1}^{(m)} \leftarrow
      V_n^{(m)}\!\left(
      \sum_{i=1}^d w_{n+1}^{i,(m)} R_{n+1}^{i,(m)}
      + w_{n+1}^{0,(m)} R_{n+1}^{0,(m)}
      \right)\)\;
      \If{\(V_{n+1}^{(m)} \le 0\)}{
        \textbf{break}
      }
    }
    \(u^{(m)} \leftarrow U\!\bigl(V_N^{(m)}\bigr)\)\;
  }

  \(\widehat{\mu} \leftarrow \frac1M \sum_{m=1}^M u^{(m)}\)\;
  \(\widehat{\psi} \leftarrow \frac1M \sum_{m=1}^M \bigl(-u^{(m)}-\eta\bigr)_+\)\;
  \(\widehat g \leftarrow \eta + \frac{1}{1-\kappa}\widehat{\psi} + \widehat{\mu}\)\;
  \(v \leftarrow \widehat g - K\)\;
  \(\bar v \leftarrow \beta \bar v + (1-\beta)v\)\;

  \(\widehat{\mathcal L} \leftarrow
  -\widehat{\mu}
  + \lambda_1 (v)_+
  + \lambda_2 \bigl((v)_+\bigr)^2\)\;

  Update \((\theta,\eta)\) by one optimizer step on \(\widehat{\mathcal L}\)\;
  $\lambda_1 \leftarrow [\lambda_1 + \rho_1\bar v]_+$,
  $\lambda_2 \leftarrow [\lambda_2 + \rho_2\bar v^2]_+$\;
}
\KwOut{Policy \(B_\theta\) and auxiliary variable \(\eta\).}
\end{algorithm}

\section{Numerical experiments under a deterministic multidimensional Black--Scholes model} \label{sec:exp_multi_bs_model}

We evaluate the proposed approach in a deterministic multidimensional Black--Scholes setting, focusing on the learned policy and its performance under various constraint specifications. 

All numerical computations are performed in Python, using \texttt{PyTorch} for the neural network-based strategies and \texttt{SciPy} for the baseline methods.

\subsection{Model specification and training setup}

We consider a market with four risky assets and one risk-free asset over a one-year horizon \(T=1\). 
The risky assets follow a multidimensional Black--Scholes model
\[
\frac{dX_t^i}{X_t^i} = \mu_i\,dt + \sum_{j=1}^d \sigma_{ij}\,dW_t^j,
\]
with initial prices \(S_0=1\), drift vector
\[
\mu = [0.09, 0.15, 0.21, 0.12],
\]
volatilities
\[
\mathbf{vol} = [0.08, 0.12, 0.15, 0.08],
\]
and correlation matrix
\[
\rho =
\begin{pmatrix}
1 & 0.2 & -0.3 & 0 \\
0.2 & 1 & 0.15 & -0.2 \\
-0.3 & 0.15 & 1 & 0.3 \\
0 & -0.2 & 0.3 & 1
\end{pmatrix}.\]
The covariance matrix is \(\Gamma = D\rho D\), with \(D=\operatorname{diag}(\mathbf{vol})\), and \(\sigma\) is obtained via Cholesky decomposition.
 The covariance matrix is \(\Gamma = D\rho D\), with \(D=\operatorname{diag}(\mathbf{vol})\), and \(\sigma\) is obtained via Cholesky decomposition. The risk-free rate is \(r=0.02\).

The initial wealth is \(V_0=100\), the DCVaR constraint level is \(K=30\), and \(\kappa=0.99\). 
The time interval is discretized into \(N\) steps (\(N=4,12,52\)).

The RNN described in Section~\ref{subsec:RNN} is trained using Monte Carlo simulation. Each epoch uses \(10^5\) simulated paths and mini-batches of size \(5{,}000\)–\(10{,}000\), corresponding to approximately \(50\)–\(100\) tail observations per batch.

Training is organized in four phases: a short warm-up phase (2–5\%), an adaptive penalty phase (60–80\%), a stabilization phase with fixed penalties, and a final full-batch phase (5–10\%) for variance reduction and optional frontier calibration.

\subsection{Training details}

The RNN cell is implemented as a feedforward neural network with two hidden layers of 50 neurons each, using ReLU activations and optional layer normalization. Parameters are initialized using Kaiming uniform initialization. Optimization is performed using AdamW with learning rates \(10^{-3}\) (network) and \(10^{-4}\) (\(\eta\)), and exponential decay.

Penalty coefficients follow the adaptive scheme of Section~\ref{sec:training_method}, with initial values in \([0.1,1]\), stepsizes \(\rho_{1,2}=0.003\), and capped updates. They typically stabilize in the range \(1\)–\(5\), sufficient to enforce the constraint. Once feasibility is reached, optimization focuses on the objective.

\subsection{Benchmark strategies}

We compare the learned policy with two classical baselines obtained via sample average approximation:
(i) a buy-and-hold strategy optimized at \(t_0\), and 
(ii) a constant-mix strategy maintaining fixed portfolio weights over time. These benchmark strategies are obtained by solving the corresponding sample average approximation (SAA) problem, using a numerical procedure analogous to that described in \cite{lelong2026}.

\subsection{Results}

We report results for three constraint configurations: long-only (LO), relative constraints (RC), and no constraints (NC), and for rebalancing frequencies \(N \in \{4,12,52\}\). Each configuration is trained independently and compared with the corresponding benchmarks.

Table~\ref{tab:results_constraints} reports the expected terminal wealth \(\E[V_N]\), the realized constraint value, and the terminal volatility \(\sigma_N=\sqrt{\Var(V_N)}\).

For the relative constraint setting, we impose the bounds 
\(\texttt{max\_bounds} = [1000,\, 0.4,\, 0.3,\, 0.4,\, 1]\) and 
\(\texttt{min\_bounds} = [-1,\, -1,\, -1,\, -1,\, -1]\). 
They are calibrated from the unconstrained buy-and-hold solution \([-0.93,\,0.58,\,0.37,\,0.34,\,0.64]\). They are chosen to limit concentration in the most profitable risky assets while preserving flexibility in the allocation. The large upper bound on the cash position effectively leaves it unconstrained from above. Overall, this specification induces binding constraints that force the policy to redistribute exposure across assets.

Across all configurations, the learned policy satisfies the DCVaR constraint and consistently outperforms naive strategies in terms of expected terminal wealth. The improvement is more pronounced in less constrained settings, reflecting the increased flexibility of the control. Under binding constraints, the network compensates by reallocating weights, resulting in more diversified and smoother portfolio profiles.

Performance increases with the number of rebalancing dates \(N\), as the policy can adapt more frequently to realized trajectories and exert finer control over the wealth dynamics. However, this effect interacts with the treatment of ruin: since the portfolio value is frozen upon hitting zero, the probability of early termination increases with \(N\), which penalizes static strategies such as buy-and-hold. This phenomenon disappears in the long-only case, where ruin cannot occur.

A deterioration of the Sharpe ratio is observed, reflecting a limitation of DCVaR-based optimization: since the objective focuses exclusively on tail risk, other distributional characteristics may be neglected whenever this improves the DCVaR criterion. This effect is visible across all constraint configurations and highlights the intrinsic limitation of single-criterion formulations, which may lead to poor mean--variance trade-offs. This naturally motivates multi-criteria extensions, such as mean--variance--CVaR models~\cite{roman2007mean}. In the present framework, a mean--variance--DCVaR formulation arises as a direct extension: the DCVaR constraint can be handled via the same Clarke-type penalization, while the variance component may be incorporated either through an additional penalty or an analogous exact-penalty construction, as in \cite{chen2017pension}.

\begin{table}[H]
\centering
\begin{tabular}{llcccccc}
\toprule
Constraint type & Strategy & $N$ & $\E[V_N]$ & Constraint value & $\sigma_N$ & Sharpe Ratio\\
\midrule
\multirow{9}{*}{LO}
 & \multirow{3}{*}{Buy-and-hold} 
   & 4  & 120.31 & 29.97 & 12.83 & 1.583 \\
 & & 12 & 120.32 & 30.17 & 12.82 & 1.585 \\
 & & 52 & 120.30 & 30.11 & 12.79 & 1.587 \\[3pt]
 & \multirow{3}{*}{Constant-mix} 
   & 4  & 120.31 & 29.98 & 12.84 & 1.582 \\
 & & 12 & 120.30 & 30.09 & 12.78 & 1.588 \\
 & & 52 & 120.24 & 30.09 & 12.78 & 1.584 \\[3pt]
 & \multirow{3}{*}{RNN} 
   & 4  & \textbf{122.03} & 29.98 & 17.48 & 1.260 \\
 & & 12 & \textbf{122.51} & 30.08 & 18.42 & 1.222 \\
 & & 52 & \textbf{122.66} & 29.99 & 18.55 & 1.222 \\[3pt]
\midrule
\multirow{9}{*}{RC}
 & \multirow{3}{*}{Buy-and-hold} 
   & 4  & 126.05 & 30.02 & 12.57 & 2.072 \\
 & & 12 & 126.13 & 30.16 & 12.68 & 2.061 \\
 & & 52 & 126.13 & 30.12 & 12.71 & 2.056 \\[3pt]
 & \multirow{3}{*}{Constant-mix} 
   & 4  & 125.98 & 29.93 & 12.53 & 2.073 \\
 & & 12 & 126.05 & 30.05 & 12.63 & 2.063 \\
 & & 52 & 125.92 & 29.84 & 12.58 & 2.060 \\[3pt]
 & \multirow{3}{*}{RNN} 
   & 4  & \textbf{126.61} & 29.94 & 13.97 & 1.905 \\
 & & 12 & \textbf{127.43} & 30.01 & 14.98 & 1.831 \\
 & & 52 & \textbf{127.75} & 30.02 & 15.61 & 1.778 \\[3pt]
\midrule
\multirow{9}{*}{NC}
 & \multirow{3}{*}{Buy-and-hold} 
   & 4  & 126.52 & 30.06 & 12.50 & 2.122 \\
 & & 12 & 126.65 & 30.12 & 12.65 & 2.107 \\
 & & 52 & 126.69 & 30.14 & 12.71 & 2.100 \\[3pt]
 & \multirow{3}{*}{Constant-mix} 
   & 4  & 126.39 & 29.92 & 12.43 & 2.123 \\
 & & 12 & 126.50 & 29.94 & 12.56 & 2.110 \\
 & & 52 & 126.63 & 30.08 & 12.68 & 2.100 \\[3pt]
 & \multirow{3}{*}{RNN} 
   & 4  & \textbf{129.70} & 30.07 & 21.09 & 1.408 \\
 & & 12 & \textbf{134.06} & 30.08 & 19.13 & 1.780 \\
 & & 52 & \textbf{136.06} & 29.43 & 21.47 & 1.680 \\
\bottomrule
\end{tabular}
\caption{Numerical results for different horizons $N \in \{4, 12, 52\}$ 
and for the three constraint types (LO: Long-Only, RC: Relative Constraint, NC: No Constraint).} \label{tab:results_constraints}
\end{table}

\section{Extension to (Re)insurance Portfolio Optimization}\label{sec:reinsurance}

We consider a dynamic portfolio optimization problem in a (re)insurance setting, where uncertainty arises from technical profit processes generated by heterogeneous lines of business (LoBs), rather than financial returns. These processes are driven by systematic risk factors and contractual cash-flow structures, and are constructed from an explicit modelling of mortality, longevity, and casualty activities; see Appendix~\ref{apendix_gompertz}.
In this context, the framework provides a basis for decision-support tools for underwriting and capital allocation under uncertainty, rather than precise forecasting models.

\subsection{Simulated lines of business and profit processes}

We consider a finite collection of insurance and reinsurance lines of business
(LoBs), indexed by $\ell\in\mathcal L$. In the insurance case studied throughout
this section, $\mathcal L$ includes mortality and longevity portfolios, whose
annual profit processes are constructed from contract-level cash-flows and best
estimate revisions, as detailed in Appendix~\ref{apendix_gompertz}. We place ourselves at calendar time \(a+n_0\), which is taken as the initial evaluation date, and simulate the portfolios dynamics over an additional horizon of \(n\) years. This formulation reflects the presence of in-force business: contracts have been underwritten from year \(a\) onward and continue to generate cash-flows and technical results at time \(a+n_0\) and beyond. The index \(n_0\) therefore represents the first simulation date, from which both the run-off of existing contracts and the effects of new underwriting decisions are accounted for.

For each
$\ell\in\{M,L\}$, where $M$ and $L$ denote mortality and longevity,
respectively, the resulting sequence
\[
\big(Y^{\ell}_{a+j}\big)_{j=n_0,\dots,n}
\]
represents the aggregate technical result of the corresponding LoB over the
simulation horizon, conditional on the evolution of a common systematic mortality factor.

The casualty component \(\ell=C\) represents short-tailed non-life risk and is
assumed independent across time and from life factors. Let \(S_{a+j}^C\) denote
the aggregate claim amount over year \(a+j\), with
\[
S_{a+j}^C \sim \Gamma(\alpha_C,1/\beta_C),
\qquad j=n_0,\dots,n.
\]
The corresponding technical result is
\[
Y_{a+j}^C
:=
(1+\theta_C)\,\frac{\alpha_C}{\beta_C}
-
S_{a+j}^C,
\]
where \(\theta_C>0\) is a proportional premium loading, so that
\((1+\theta_C)\alpha_C/\beta_C\) represents the premium income. In particular,
\(\mathbb E[Y_{a+j}^C]=\theta_C\,\alpha_C/\beta_C>0\).

This stylised specification provides a tractable source of diversifying risk.
Throughout, the joint law of the LoB profit processes is obtained by Monte Carlo
simulation and treated as given in the optimization problem.

\subsection{Dependency structure and admissible controls}

In the (re)insurance setting, the profit processes
\(\big(Y^{\ell}_{a+j}\big)\) exhibit both intertemporal and cross-line
dependence. Mortality and longevity results are jointly driven by a common
systematic factor, inducing correlated outcomes across time and lines of
business, with heterogeneous sensitivities due to contractual cash-flow
structures. Diversification therefore arises endogenously from the interaction
between activities.

Admissible controls fundamentally differ from classical asset allocation.
Decisions \(\alpha_i=(\alpha_i^\ell)_{\ell\in\mathcal L}\) represent
underwriting volumes or exposures rather than portfolio weights, and are not
subject to a self-financing constraint. In life insurance segments, these
decisions have persistent effects: a control at time \(i\) impacts the entire
future stream \(\big(X^{\ell}_{i,d}\big)_{d\ge1}\) for
\(\ell\in\{M,L\}\) (see Appendix~\ref{appendix_gompertz_LoB}),
reflecting the long-term nature of contracts, in contrast with the
local-in-time effects of financial controls.

The optimization is formulated over a finite horizon with a small number of
decision dates (typically \(N\le 10\)).

\subsection{Problem formulation}

All sources of randomness are defined on a filtered probability space
\((\Omega,\mathbb F,\mathbb P)\), with
\(\mathbb F=(\cF_i)_{0\le i\le T}\) representing the information available at
calendar time \(i\). A deterministic discount factor \(v=(1+r)^{-1}\) is assumed.

Let \(\mathcal L=\{M,L,C\}\). For each underwriting year \(i\) and
\(\ell\in\mathcal L\), the profit streams
\(\big(X^{\ell}_{i,d}\big)_{d\ge1}\) are constructed as in
Appendix~\ref{apendix_gompertz}, capturing long-term contractual effects in
life segments and short-tailed independent dynamics in the casualty segment.

At each time \(i\), the insurer selects a control
\(\alpha_i=(\alpha_i^\ell)_{\ell\in\mathcal L}\), where
\((\alpha_i)_{i=n_0,\dots,n}\) is \(\mathbb F\)-predictable and takes values in a prescribed admissible set $\mathcal A$, reflecting operational, regulatory, or capacity constraints.

Given a control strategy and a fixed in-force portfolio \(\alpha_{IF}\), the
aggregate profit of LoB \(\ell\) at calendar time \(a+j\) is
\[
Y^{\ell}_{a+j}
=
\sum_{k=a}^{a+n_0-1}
\alpha_{IF,k}\,X^{\ell}_{k,\,a+j-k}
+
\sum_{i=a+n_0}^{a+j}
\alpha_i^{\ell}\,X^{\ell}_{i,\,a+j-i},
\qquad j=n_0,\dots,n,
\]
where the first term represents the run-off of in-force business and the second
the contribution of new underwriting decisions.

The resulting processes are generally non-Markovian and non time-separable, as
controls induce persistent effects on future cash-flows.

The objective is to maximize the discounted cumulative profit
\[
\mathbb{E}\!\left[
\sum_{\ell\in\mathcal L}
\sum_{j=n_0}^{n}
v^{\,j+1-n_0}\,
Y^{\ell}_{a+j}
\;\middle|\;
\cF_{n_0-1}
\right],
\]
subject to the global DCVaR constraint
\[
\mathrm{DCVaR}_{\kappa}\!\left(
-\sum_{\ell\in\mathcal L}
\sum_{j=n_0}^{n}
v^{\,j+1-n_0}\,
Y^{\ell}_{a+j}
\;\middle|\;
\cF_{n_0-1}
\right)
\le K.
\]

This defines a multi-period insurance portfolio control problem under DCVaR
constraints, where decisions are taken sequentially based on available
information and the evolution of systematic risk factors.

\subsection{Policy parameterization}

We consider the construction of an admissible \(\mathbb F\)-predictable control
sequence \((\alpha_i)_{i=n_0,\dots,n}\) governing underwriting decisions across
lines of business. In contrast with financial portfolio models, the controlled
profit processes are neither Markovian nor self-financing, as their evolution
depends on the full history of past decisions and realised claims. As a
consequence, optimal controls cannot, in general, be represented as functions of
a finite-dimensional state, which precludes the direct use of dynamic
programming methods.

We approximate admissible strategies by a parametric, history-dependent policy
\[
\alpha_i = A_\theta(\mathcal I_i),
\qquad i=n_0,\dots,n,
\]
where \(\mathcal I_i\) denotes the information available at time \(i\) and
\(\theta\) the policy parameters.

In practice, \(A_\theta\) is implemented as a recurrent neural network (RNN),
as described in Subsection~\ref{subsec:RNN}, with internal dynamics
\[
h_i = \Phi_\theta(h_{i-1}, x_i),
\qquad
\alpha_i = \phi_\theta(h_i),
\]
where \(h_i\) is a latent state summarizing past information and \(x_i\) the
current observable features.

The transition map \(\Phi_\theta\) is implemented as a feedforward neural network,
applied at each time step, optionally equipped with residual (skip) connections.

This implicit state augmentation allows the policy
to capture nonlinear and long-range dependencies induced by insurance
liabilities.

Admissibility is enforced by construction through projection onto a convex set \(\mathcal A\), typically encoding long-only or box-type constraints on underwriting volumes.

\subsection{Training methodology}

The numerical resolution follows the training framework introduced in
Section~\ref{sec:training_method}. In particular, the global DCVaR constraint is
handled via an exact penalty approach, leading to the penalized objective
\begin{equation}
\label{eq:penalized_objective_insurance}
\mathcal{L}(\btheta,\eta)
=
f(\btheta)
+
\lambda_{1}\big(g(\btheta,\eta)-K\big)_+
+
\lambda_{2}\big(g(\btheta,\eta)-K\big)_+^{2},
\end{equation}
where \(\eta\) is the auxiliary variable associated with the DCVaR functional.

The parameters \((\btheta,\eta)\) are updated jointly using stochastic gradient
steps. In addition, \(\eta\) is refined through a few inner optimisation steps,
with \(\btheta\) fixed, to minimise \(g(\btheta,\eta)\) with respect to the
quantile parameter. This two-timescale update allows \(\eta\) to track the
relevant risk quantile more accurately and improves numerical stability in the
multi-period setting.

\subsection{Numerical Experiments on an Illustrative Example}
\label{sec:numerical_experiment}

We report numerical experiments on a stylised insurance portfolio to illustrate
the proposed multi-period optimisation framework. We describe the simulation
model, calibration procedure, and evaluate out-of-sample performance under a
global DCVaR constraint.

\subsubsection{Simulation model and parameter specification}
\label{app:numexp_simulation_model}

\paragraph{Model specification.}
Systematic mortality and longevity risks (see Appendix~\ref{apendix_gompertz})
are driven by two correlated square-root diffusions (CIR processes),
discretised on a yearly time grid. In the numerical experiments, both factors
share parameters \((g_\ell,b_\ell,\sigma_\ell)=(0.30,0.30,0.20)\), with
correlation \(\rho=0.25\) and initial levels \(Z_0^M=Z_0^L=1\). The parameter
choice is stylised and aims at producing realistic variability rather than
precise calibration. Simulation is performed using a positivity-preserving Euler
scheme with full truncation; see
\cite{lord2010comparison}.

A short-tailed casualty LoB is added via i.i.d.\ annual aggregate claims
\(S^C_{a+j}\sim\Gamma(\alpha_C,1/\beta_C)\) with
\((\alpha_C,\beta_C)=(5,1)\), independent of life factors, providing a
diversifying non-life component.

\paragraph{Premium calibration and normalisation.}
For mortality and longevity portfolios, contract parameters are calibrated under
actuarial fairness, equating discounted expected benefits and premiums at
inception. Commercial premiums are then obtained by applying a uniform loading
factor \(\theta=0.10\).

Exposures are subsequently normalised so that one unit of underwriting volume
\(\alpha^\ell=1\) yields the same expected economic margin across LoBs. In the
numerical experiments, this corresponds approximately to an expected discounted
premium inflow of \(11\) and an expected discounted benefit outflow of \(10\)
per unit of exposure, the difference being induced by the common premium
loading. This normalisation ensures that optimal allocations reflect differences in risk and dependence structures rather than scale effects.

\subsubsection{Learning parameters and training algorithm for the neural policy}
\label{app:nn_training}

The admissible control is parameterized by a shared recurrent policy
$\pi_{\btheta}$, which outputs, at each decision date
$s=0,\dots,n_{\mathrm{ctrl}}-1$ (with $n_{\mathrm{ctrl}}=5$), the underwriting decisions
$\alpha_s=(\alpha_s^M,\alpha_s^L,\alpha_s^C)\in\R^{3}$.

The policy is implemented as a recurrent neural network (RNN), whose cell is modeled by a gated recurrent unit (GRU) with hidden size $16$. The recurrent output is then processed by a feedforward head consisting of a residual MLP, followed by a final linear projection onto the control space.

At each step, the network takes as input a feature vector combining time features, the current aggregate state, the previous decision $\alpha_{s-1}$, and cohort-level state variables $(N_s^{M},N_s^{L})\in\R^{A}$ summarizing in-force exposures by underwriting year, optionally augmented with a cohort mask. The resulting input dimension is $d=5+2A$ (or $d=5+3A$), leading to a compact parametrization with $\mathcal{O}(H^2+Hd)$ parameters.

Finally, a smooth projection layer is used to enforce box constraints, ensuring differentiability and compatibility with gradient-based optimization.

\subsubsection{Numerical results}
\label{subsubsec:numerical_results}

We consider six configurations obtained by combining three constraint regimes on the controls, namely \emph{long-only} (LO), \emph{constant bounds} (CSTB), and \emph{time-dependent bounds} (TDB), with two in-force specifications: IF00 with $(\alpha_{IF}^M,\alpha_{IF}^L)=(0,0)$ and IF11 with $(\alpha_{IF}^M,\alpha_{IF}^L)=(1,1)$. Constraints apply componentwise to $(\alpha_M,\alpha_L,\alpha_C)$, while the casualty in-force component is set to zero due to its absence of intertemporal effects. Additional details are provided in Appendix~\ref{app:extended_nr}.

All policies are trained on $5\times10^{5}$ simulated trajectories and evaluated on an independent test set. The penalised objective is optimized with $\lambda_1,\lambda_2\in[3,10]$, using AdamW over $400$ to $500$ epochs with mini-batches of size $2.5\times10^4$, following the training strategy of Section~\ref{sec:exp_multi_bs_model}.

We compare the recurrent policy with two static benchmarks: a constant strategy (Const) and a fixed time-dependent profile (FT), both obtained via constrained SAA solved by SLSQP~\cite{lelong2026}. Performance is measured by expected discounted profit and feasibility under a DCVaR constraint with $K=30$ and $\kappa=0.99$.

To mitigate the slight upward bias induced by training on a fixed Monte Carlo sample, we apply a frontier calibration by training under a tightened constraint $K-\delta$. Benchmarks are then recalibrated at the attained risk level, ensuring a fair comparison.

\begin{table}[!h]
\centering
\small
\setlength{\tabcolsep}{5pt}
\begin{tabular}{ll l rrrr}
\toprule
IF & Bounds & Policy 
& $\mathbb{E}[R]$ 
& $\mathrm{Std}(R)$ 
& $\mathrm{DCVaR}_{0.99}(-R)$ 
& $\mathrm{VaR}_{0.99}(-R)$ \\
\midrule
\multirow[c]{9}{*}{IF00}
& \multirow[c]{3}{*}{LO}
& NN    & \textbf{98.280} & 17.718 & 30.030 & -72.799 \\
&       & FT    & 71.671 & 11.385 & 30.128 & -45.425 \\
&       & Const & 57.422 & 11.380 & 30.150 & -31.152 \\
\cmidrule(lr){2-7}
& \multirow[c]{3}{*}{CSTB}
& NN    & \textbf{81.861} & 11.912 & 30.080 & -56.306 \\
&       & FT    & 66.062 & 11.291 & 30.224 & -39.746 \\
&       & Const & 57.422 & 11.379 & 30.150 & -31.150 \\
\cmidrule(lr){2-7}
& \multirow[c]{3}{*}{TDB}
& NN    & \textbf{60.312} & 11.276 & 29.864 & -34.265 \\
&       & FT    & 60.344 & 11.286 & 29.898 & -34.254 \\
&       & Const & 28.433 & 10.196 & 28.851 &  -3.544 \\
\midrule
\multirow[c]{9}{*}{IF11}
& \multirow[c]{3}{*}{LO}
& NN    & \textbf{92.078} & 25.529 & 30.031 & -66.661 \\
&       & FT    & 24.327 & 11.095 & 29.770 &   1.651 \\
&       & Const & 16.703 & 11.173 & 29.871 &   9.336 \\
\cmidrule(lr){2-7}
& \multirow[c]{3}{*}{CSTB}
& NN    & \textbf{53.707} & 13.191 & 30.006 & -29.473 \\
&       & FT    & 20.140 & 11.003 & 29.829 &   5.856 \\
&       & Const & 16.316 & 11.033 & 29.916 &   9.733 \\
\cmidrule(lr){2-7}
& \multirow[c]{3}{*}{TDB}
& NN    & \textbf{41.242} &  9.776 & 30.198 & -17.098 \\
&       & FT    & 20.724 & 11.008 & 29.809 &   5.234 \\
&       & Const & 15.957 & 10.963 & 29.935 &  10.092 \\
\bottomrule
\end{tabular}
\caption{Out-of-sample performance (evaluation set) under both in-force configurations.
Reported are the mean and standard deviation of the reward,
the empirical $\mathrm{DCVaR}_{0.99}$, and the $\mathrm{VaR}_{0.99}$ of the loss.}
\label{tab:results_eval_summary}
\end{table}

\FloatBarrier

\medskip

\paragraph{Global analysis.}

\begin{figure}[h!]
\centering
\begin{tikzpicture}
\begin{axis}[
    width=13cm,height=7cm,
    axis lines=left,
    xlabel={$y$}, ylabel={density},
    xmin=-9, xmax=3,
    ymin=0,
    ymax=0.5,
    legend style={at={(0.98,0.98)},anchor=north east},
    samples=250,
    domain=-9:3
]

\def\muA{-1}
\def\muB{-4}
\def\s{1}
\def\z{2.3263478740408408}

\addplot[thick] {1/(\s*sqrt(2*pi))*exp(-((x-\muA)^2)/(2*\s^2))};
\addlegendentry{$-R$}

\addplot[thick, dashed] {1/(\s*sqrt(2*pi))*exp(-((x-\muB)^2)/(2*\s^2))};
\addlegendentry{$-R_{\alpha^*}$}

\addplot[very thick] coordinates {(\muA,0) (\muA,0.45)};
\node[rotate=90, anchor=south] at (axis cs:\muA,0.24)
{$\mathbb{E}[-R_{\alpha}]$};

\addplot[very thick] coordinates {(\muB,0) (\muB,0.45)};
\node[rotate=90, anchor=south] at (axis cs:\muB,0.24)
{$\mathbb{E}[-R_{\alpha^*}]$};

\pgfmathsetmacro{\varA}{\muA + \z*\s}
\pgfmathsetmacro{\varB}{\muB + \z*\s}

\addplot[densely dotted, very thick] coordinates {(\varA,0) (\varA,0.45)};
\node[rotate=90, anchor=south] at (axis cs:\varA,0.24)
{$\mathrm{VaR}_{0.99}(-R_{\alpha})$};

\addplot[densely dotted, very thick] coordinates {(\varB,0) (\varB,0.45)};
\node[rotate=90, anchor=south] at (axis cs:\varB,0.24)
{$\mathrm{VaR}_{0.99}(-R_{\alpha^*})$};

\draw[->, thick]
(axis cs:\muA,0.44) -- (axis cs:\muB,0.44)
node[midway, above] {$\text{Optimization-induced mean shift}$};

\end{axis}
\end{tikzpicture}
\caption{Illustration of the optimization-induced shift of the loss distribution.
The displacement of $\mathrm{VaR}_{0.99}(-R)$ is a direct consequence of the leftward translation of $-R$ under the optimal allocation.}
\label{fig:var_shift_loss}
\end{figure}
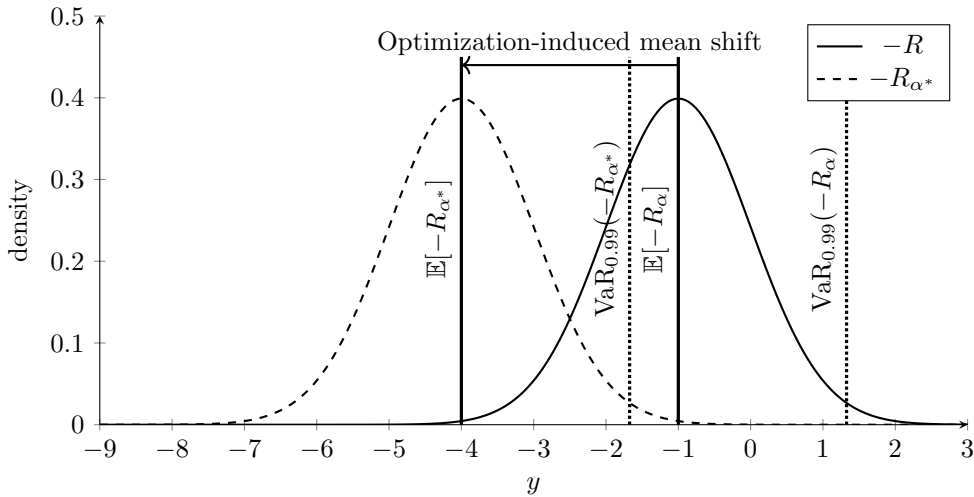

Across most configurations, neural policies outperform static baselines in expected reward while maintaining comparable or lower DCVaR levels (Table~\ref{tab:results_eval_summary}). The improvement is particularly pronounced under loose constraints, where dynamic reallocation yields significant gains without increasing tail risk. This is reflected in a leftward shift of the loss distribution, with $\mathrm{VaR}_{0.99}(-R)$ becoming more negative, indicating positive profit even in extreme scenarios (Figure~\ref{fig:var_shift_loss}).

As constraints become more restrictive, differences between policies decrease and may vanish when the admissible set is binding, as observed in the IF00 TDB case. In this regime, the in-force portfolio dominates the risk dynamics while contributing little to expected return, which limits the benefit of dynamic control. Neural policies nevertheless remain competitive and typically achieve the most favorable trade-off between performance and tail risk.

As in the financial setting, variance may deteriorate since the objective focuses exclusively on DCVaR. Extensions incorporating variance, as discussed earlier, can be implemented within the same framework.

Overall, these results highlight the benefit of adaptive and history-dependent controls in the multi-period insurance setting. The recurrent architecture captures the intertemporal effects of underwriting decisions and mitigates the risk contribution of the in-force component, which static strategies fail to reproduce.

A detailed description of the constraint configurations, together with supplementary numerical results and discussion, is provided in Appendix~\ref{app:extended_nr}.

\subsubsection{Discussion}

While the proposed framework highlights the benefits of adaptive multi-period control under a global DCVaR constraint, several extensions are needed to better align with regulatory and operational practice. In particular, risk constraints should be enforced not only at the terminal horizon but also at intermediate dates, since the numerical results reveal significant heterogeneity in controls over time, whereas regulatory metrics are typically defined on a one-year basis.

Incorporating an explicit cost of capital would further enhance economic realism. As regulatory capital requirements are closely linked to deviation-based risk measures, DCVaR provides a natural proxy for capital intensity. Including it as a running cost would allow the model to capture the trade-off between profitability and capital consumption dynamically.

More broadly, these results underline the limitations of sequential single-period approaches. In life insurance portfolios, early underwriting decisions have persistent effects on future risk profiles, which cannot be captured by myopic or rolling optimizations. A genuinely multi-period framework is therefore required to control the intertemporal accumulation of tail risk.

\section{Conclusion}

We studied a discrete-time multi-period portfolio optimization problem under an explicit Deviation Conditional Value-at-Risk (DCVaR) constraint, leading to a time-inconsistent precommitment formulation in which expected performance is maximized subject to a global tail-risk bound.

To address the computational challenges induced by path dependence and high-dimensional state variables, we proposed a recurrent neural-network-based approximation of the optimal policy. The method enforces the DCVaR constraint directly within a simulation-based framework via an exact penalization scheme, avoiding dynamic programming recursions and restrictive structural assumptions.

Across both financial and insurance applications, the results show that adaptive, history-dependent policies improve expected performance while satisfying the prescribed risk level. Gains are most pronounced when the admissible control space remains sufficiently flexible, whereas under binding constraints dynamic and static strategies converge.

A central feature of this work is the explicit treatment of DCVaR as a constraint, preserving the target risk level $K$ as a structural parameter of the problem. In contrast to approaches that restore time consistency through utility reformulations or rely on complete-market martingale techniques, we deliberately retain the precommitment structure of the mean--DCVaR problem and enforce the constraint numerically via an exact penalization scheme, ensuring a clear interpretation of feasibility.

While much of the literature focuses on mean--CVaR formulations, this work emphasizes the deviation-based measure DCVaR in a genuinely multi-period and explicitly constrained setting. As discussed earlier, DCVaR provides a natural measure of dispersion around the expectation and is particularly relevant in contexts where capital intensity is a primary concern, notably in insurance applications.

More broadly, the proposed framework provides a scalable approach to multi-period risk-constrained optimization under complex and path-dependent dynamics. It extends deviation-based formulations beyond classical mean--CVaR settings and enables the treatment of high-dimensional portfolio problems without relying on restrictive structural assumptions.

\appendix
\section{Stochastic Gompertz--Makeham Mortality Model for Line-of-Business Modelling} \label{apendix_gompertz}

This appendix specifies the simulation framework for a life insurance portfolio
combining mortality and longevity products. Mortality dynamics are modeled using
a stochastic Gompertz--Makeham specification, capturing age-dependent mortality
patterns and systematic longevity risk.

We define the contractual cash flows for mortality and longevity products and
derive the associated line-of-business profit processes. These constructions
provide the basis for the simulation of multi-period portfolio outcomes used in
the numerical experiments.

\subsection{Mortality intensity and death probabilities}

We adopt a Gompertz--Makeham specification augmented by a stochastic mortality
factor, following the actuarial literature on factor-based mortality modelling;
see, e.g.,~\cite{milevsky2001mortality,biffis2005affine,cairns2006two,
luciano2008mortality}. The baseline force of mortality at age $x$ is given by
\[
\mu_0(x) = a + \beta e^{c x}, \qquad a,\beta,c>0,
\]
where the constant term $a$ represents age-independent mortality, while the
exponential component $\beta e^{c x}$ captures the Gompertz senescence effect.

Systematic mortality risk is modelled through a strictly positive stochastic
intensity factor $(Z_t)_{t\ge0}$, assumed to follow a Cox--Ingersoll--Ross (CIR)
square-root diffusion~\cite{cox1985theory},
\[
dZ_t = (g - b Z_t)\,dt + \sigma \sqrt{Z_t}\,dB_t, \qquad Z_0>0,
\]
where $(B_t)_{t\ge0}$ denotes a standard Brownian motion and $g,b,\sigma>0$ are
parameters chosen so as to ensure positivity of the process. The parameter $b$ governs the speed of mean reversion toward the
long-run level $g/b$, while $\sigma$ controls the instantaneous volatility of
the intensity. The CIR dynamics
belong to the class of affine stochastic intensity models and are widely used in
actuarial and financial applications.

The resulting stochastic force of mortality for an individual aged $x$ at
calendar time $t$ is defined multiplicatively as
\[
\mu(x,t) = Z_t\,\mu_0(x).
\]

Given this continuous-time formulation, the one-year conditional probability of
death for an individual aged $x$ at time $t$ is given by
\[
q_x(t)
=
1 - \exp\!\left(
-\int_0^1 Z_{t+s}\,\mu_0(x+s)\,ds
\right).
\]

For simulation purposes, and to remain consistent with the discrete-time
structure of the numerical framework, we approximate the systematic factor by a
piecewise-constant process over yearly intervals, setting $Z_{t+s}\approx Z_t$
for $s\in[0,1)$. This yields the approximation
\[
q_x(t)
\approx
1-\exp\!\left(
- Z_t \int_0^1 \mu_0(x+s)\,ds
\right)
=
1-\exp\!\left(
- Z_t\left[
a+\frac{\beta}{c}\big(e^{c(x+1)}-e^{cx}\big)
\right]
\right).
\]

This specification separates idiosyncratic mortality risk, arising from
individual death events, from systematic longevity risk driven by a common
stochastic factor. It yields a tractable model for multi-period simulation of
mortality and longevity dynamics.

\subsection{Cash--flow modeling of mortality and longevity portfolios}

We consider two types of homogeneous life insurance portfolios indexed by
$\ell\in\mathcal L:=\{M,L\}$, where $\ell=M$ denotes a mortality portfolio and
$\ell=L$ a longevity portfolio. Each portfolio consists of $N_0^{\ell}$ policyholders,
all aged $x^{\ell}$ at time $t=0$, and is observed over a finite time horizon
$T\in\mathbb{N}$. A deterministic discount rate $r\ge 0$ is assumed, and we denote
$v=(1+r)^{-1}$.

For each $\ell\in\mathcal L$, let $N_t^{\ell}$ denote the number of surviving
policyholders at the beginning of year $t$ and $D_t^{\ell}$ the number of deaths
occurring over $[t,t+1)$. The corresponding one--year conditional probability of
death $q_x^{\ell}(t)$ is induced by a line--specific, possibly correlated,
specification of the stochastic mortality intensity introduced in the previous
subsection.

Conditionally on the path of the systematic mortality factors over $[t,t+1)$,
deaths are assumed independent and satisfy
\[
D_t^{\ell} \mid (Z^{\ell},N_t^{\ell}) \sim \mathrm{Binomial}\big(N_t^{\ell},q_x^{\ell}(t)\big),
\qquad \ell\in\mathcal L.
\]

\paragraph{Portfolio cash--flows.}
The net portfolio cash--flow at time $t$ is denoted by $\CF_t^{\ell}$ and is
defined as follows.

\begin{itemize}
\item \emph{Mortality portfolio $(\ell=M)$.}  
Premiums of constant amount $P^{M}$ are paid annually in advance while the
policyholder is alive, and a death benefit $C$ is paid at the end of the year of
death. The net cash--flow is therefore given by
\[
\CF^{M}_t = P^{M} N_t^{M} - C D_t^{M} .
\]

\item \emph{Longevity portfolio $(\ell=L)$.}  
We consider an immediate life annuity contract with a single premium paid at inception. A lump--sum premium $P^{L}$ is paid at time $t=0$, and surviving policyholders receive a life--contingent annuity payment of fixed amount $R$ at the end of each year until death. The corresponding net cash--flow is given by
\[
\CF^{L}_t =
\begin{cases}
\;\;\;\; P^{L} N_0^{L}, & t=0,\\[4pt]
- R N_t^{L}, & t=1,\dots,T.
\end{cases}
\]

\end{itemize}

\paragraph{Actuarial present value and fair premium.}
For each $\ell\in\mathcal L$, the actuarial present value (APV) of the portfolio
is defined as the discounted expectation of future net cash--flows,
\[
\APV^{\ell} := \sum_{t=0}^{T-1} v^{t+1}\,\mathbb{E}\!\left[\CF^{\ell}_t\right].
\]

The actuarially fair (pure) premium $P^{\ell,\mathrm{pure}}$ is obtained by
imposing the fair--pricing condition $\APV^{\ell}=0$. This yields an explicit
representation of $P^{\ell,\mathrm{pure}}$ as a ratio of discounted expected
benefit outflows to discounted expected premium inflows, which depends
exclusively on the survival expectations generated by the stochastic mortality
model.

\paragraph{Best estimate liability and premium loading.}

The best estimate liability (BEL) associated with portfolio $\ell$ is defined as
the expected present value of future benefit payments net of premium inflows,
\[
\BEL^{\ell} := -\,\APV^{\ell}.
\]
In particular, $\BEL^{\ell}=0$ under actuarially fair premiums.

In practice, a proportional loading factor $\theta>0$ is applied to the pure
premium, yielding
\[
P^{\ell} = (1+\theta)\,P^{\ell,\mathrm{pure}}.
\]
The resulting expected present value is strictly positive and represents the
insurer’s expected margin.

\subsection{Line of business modeling} \label{appendix_gompertz_LoB}

We aggregate the contract-level profit processes into a line-of-business
framework suitable for dynamic optimisation.

We adopt the standard actuarial $(a,d)$ indexing, where $a$ denotes the
underwriting year and $d$ the development year, so that calendar time is
$t=a+d$.

Each calendar year, a volume of new mortality and longevity business is
underwritten. At inception, the profit corresponds to the expected present value
of future cash flows. In subsequent years, profits arise from experience
deviations and revisions of best estimates, in line with market-consistent
valuation principles (see, e.g.,~\cite{bauer2010solvency}).

For a given line of business, cash flows at development year $d$ correspond to
the contractual exposure over $[a+d,a+d+1)$ and are settled at time $a+d+1$.
Thus, $CF_{a,d}$ is a real-valued random variable measurable with respect to
$\cF_{a+d+1}$.

\paragraph{Best estimate of future cash-flows.}

The best estimate of remaining future cash flows at calendar time $a+d$, for $d \ge 0$, is defined by
\[
\mathrm{BE}_{a,d}
:=
\mathbb{E}\!\left[
\sum_{i=d}^{N_{\max}}
v^{\,i-d+1}\, CF_{a,i}
\;\middle|\;
\cF_{a+d}
\right],
\]
where $N_{\max}$ denotes the maximal development horizon.

\paragraph{Initial recognition and first annual result.}

Contracts are assumed to be issued before underwriting year $a$, so that initial terms are $\cF_a$-measurable. At the first reporting date $a+1$, the first recognised amount after inception is
\[
X_{a,1} := CF_{a,0} + \mathrm{BE}_{a,1}.
\]
It represents the economic value of the block at that date, combining realised experience over the first coverage year with the updated estimate of remaining future cash-flows.

\paragraph{Subsequent annual profit.}
For subsequent years, the annual profit attributed to underwriting year $a$ and
development year $d$ corresponds to the profit generated over the calendar year
$[a+d-1,a+d)$ and is defined, for $d \ge 2$, by
\begin{align}
X_{a,d}
&:= CF_{a,d-1}
+\mathrm{BE}_{a,d}
-\frac{1}{v}\mathrm{BE}_{a,d-1}, \label{eq:Xd_definition}\\
&=
\underbrace{
\Big(
CF_{a,d-1}
-
\mathbb{E}\!\left[CF_{a,d-1}\mid\cF_{a+d-1}\right]
\Big)
}_{\text{realised experience deviation}}
\;+\;
\underbrace{
\Big(
\mathrm{BE}_{a,d}
-
\mathbb{E}\!\left[\mathrm{BE}_{a,d}\mid\cF_{a+d-1}\right]
\Big)
}_{\text{revision of future cash--flow estimates}}.
\label{eq:Xd_decomposition}
\end{align}

For $d \ge 2$, this decomposition highlights that the annual profit arises from both realised experience deviations and revisions of the best estimate of remaining future cash--flows. Moreover, by construction, $\mathbb{E}\!\left[X_{a,d}\mid\cF_{a+d-1}\right]=0$.

Furthermore, $X_{a,d}$ is an $\cF_{a+d}$-measurable real-valued random variable
for all $d\ge1$.

\paragraph{Large portfolio approximation.}

We adopt a classical large homogeneous portfolio approximation, whereby
idiosyncratic mortality risk is diversified away via a law of large numbers
argument; see, e.g.,~\cite{biffis2006bidimensional,dahl2004stochastic}. 
Conditionally on the filtration $\mathbb F$ generated by the systematic mortality factor, individual death events are replaced by their expectations, and the survivor process satisfies
\[
\mathbb{E}[D_t \mid \cF_t] = N_t q_t,
\qquad
N_{t+1} = N_t(1-q_t),
\]
where $q_t$ denotes the one-year death probability implied by the stochastic mortality intensity.

Under this approximation, portfolio cash--flows and best estimate liabilities
become \emph{conditionally deterministic} functions of the underlying mortality
factor $(Z_t)_{t\ge0}$. This eliminates the need for nested Monte Carlo
simulations while preserving the economically relevant systematic longevity
risk. The impact of this factor differs across lines of business, as death
events may either trigger or terminate contractual cash--flows depending on the
nature of the underlying contracts.

\paragraph{Numerical approximation of the best estimate.}

To simulate the annual profit process $(X_{a,d})_{d\ge1}$, we generate paths of the systematic mortality factor $(Z_t)_{t\ge0}$ under the stochastic Gompertz--Makeham specification. Under the large portfolio approximation, the survivor process $(N_t)$ and cash-flows $(CF_t)$ are deterministic along each path.

For mortality and longevity portfolios, the best estimate $\mathrm{BE}_{a,d}$ is the discounted conditional expectation of future cash-flows. Since cash-flows are proportional to the surviving exposure, their evaluation reduces to conditional expectations of the survivor process. For $s<a+d\le t$,
\[
\mathbb{E}[CF_{a,t}\mid\cF_s] \propto \mathbb{E}[N_t \mid \cF_s],
\qquad
\mathbb{E}[N_t \mid \cF_s]
=
N_s\,
\mathbb{E}\!\left[
\prod_{k=s}^{t-1}\exp(-K_{x+k-s} Z_k)
\;\middle|\;
Z_s
\right],
\]
where $K_x := \int_0^1 \mu_0(x+u)\,du$.

By the Markov property of $(Z_t)$, these expectations depend only on $(Z_s,h)$ with $h=t-s$, which motivates the functions
\[
f_{h,x}(z)
=
\mathbb{E}\!\left[
\exp\!\left(-\sum_{k=0}^{h-1} K_{x+k} Z_{s+k}\right)
\;\middle|\;
Z_s=z
\right].
\]
Once these functions are known, the best estimate $\mathrm{BE}_{a,d}$ can be expressed as a function of the current systematic state, with $h=d$.

The functions $(f_{h,x})$ satisfy a backward recursion
\[
f_{h,x}(z)
=
\mathbb{E}\!\left[
\exp(-K_x z)\, f_{h-1,x}(Z_{s+1})
\;\middle|\;
Z_s=z
\right],
\qquad f_0(z)=1.
\]

Since $(Z_t)$ is a CIR process, it is affine and its conditional Laplace transform admits an exponential-affine form (see~\cite{alfonsi2005discretization,alfonsi2010high}). As a result, the recursion preserves this structure and yields
\[
f_{h,x}(z)=\exp(A_{h,x}-B_{h,x}z),
\]
where $(A_{h,x},B_{h,x})$ are computed by a deterministic backward recursion. This provides an exact evaluation of conditional survival expectations without nested Monte Carlo.

\paragraph{Total line of business modelling.}

We consider the aggregate modelling of a life insurance line of business combining mortality and longevity activities. The initial evaluation date is set to calendar time $a+n_0$, and portfolio dynamics are simulated over the horizon $[a+n_0,\,a+n]$ using $M$ Monte Carlo scenarios.

For each calendar year $a+j$, $j=n_0,\dots,n$, and each line of business $\ell\in\mathcal L$, the total result is defined by
\[
Y^{\ell}_{a+j}
:=
\sum_{k=a}^{a+n_0-1}
\alpha_{IF,k}\,X^{\ell}_{k,\,a+j-k}
\;+\;
\sum_{i=a+n_0}^{a+j}
\alpha_i^{\ell}\,X^{\ell}_{i,\,a+j-i}.
\]
Here, $X^{\ell}_{k,d}$ denotes the annual profit of line of business $\ell$ associated with underwriting year $k$ and development year $d$. The coefficients $\alpha_{IF,k}$ represent fixed legacy exposures, while $(\alpha_i^{\ell})_{i\ge a+n_0}$ denote management decisions taken at calendar time $i$, based on the information available up to time $i-1$, so that the control process is $\mathbb F$-predictable.

The first term corresponds to the run-off of the in-force portfolio at time $a+n_0$, while the second term captures the contribution of new underwriting and portfolio rebalancing decisions. This decomposition distinguishes between legacy exposure and actively managed risk.

\subsection{Extended Numerical Results} \label{app:extended_nr}

This appendix reports additional numerical results and implementation details complementing Section~\ref{sec:numerical_experiment}. In particular, we provide explicit admissibility constraints and extended performance tables omitted from the main text.

\paragraph{Impact of in-force and new business composition.}

Table~\ref{tab:ifonly_comparison} highlights the sensitivity of the portfolio risk profile to the composition of the in-force book. The longevity-only configuration $(0,1,0)$ exhibits significantly lower tail risk, as measured by both DCVaR and extreme quantiles. Introducing mortality exposure, either alone $(1,0,0)$ or jointly with longevity $(1,1,0)$, leads to a substantial increase in tail risk, with limited differences between these two cases. Over the one-year horizon considered here, the casualty component $C$ has no material impact and is therefore set to zero in the subsequent analysis.

Table~\ref{tab:nbonly_comparison} reports the corresponding results for new business allocations under a time-homogeneous policy, where a constant allocation $\alpha_{\mathrm{NB}}$ is applied at each decision date over the five-year horizon. In this setting, the configuration $(0,0,1)$ exhibits markedly higher volatility and tail risk, reflecting the accumulation of risk over time. In contrast, diversified allocations such as $(1,1,0)$ or $(1,1,1)$ achieve higher expected returns with more controlled tail risk, illustrating the benefits of combining multiple business lines.

\begin{table}[!h]
\centering
\small
\setlength{\tabcolsep}{6pt}
\begin{tabular}{lrrrr}
\toprule
$\alpha_{\mathrm{IF}}$ &
$\mathbb{E}[R]$ &
$\mathrm{Std}(R)$ &
$\mathrm{DCVaR}_{\mathrm{emp}}$ &
$\mathrm{VaR}_{0.99}(-R)$ \\
\midrule
$(0,1,0)$ & 0.008 & 4.546 & 11.780 & 10.383 \\
$(1,0,0)$ & 0.082 & 8.474 & 24.054 & 20.710 \\
$(1,1,0)$ & 0.089 & 8.607 & 24.028 & 20.771 \\
\bottomrule
\end{tabular}
\caption{Portfolio performance under fixed controls for different in-force compositions
$\alpha_{\mathrm{IF}}$ on the training set.}
\label{tab:ifonly_comparison}
\end{table}

\begin{table}[!h]
\centering
\small
\setlength{\tabcolsep}{6pt}
\begin{tabular}{lrrrr}
\toprule
$\alpha_{\mathrm{NB}}$ &
$\mathbb{E}[R]$ &
$\mathrm{Std}(R)$ &
$\mathrm{DCVaR}_{\mathrm{emp}}$ &
$\mathrm{VaR}_{0.99}(-R)$ \\
\midrule
$(1,0,0)$ &  4.810 &  1.364 &  4.057 &  -1.322 \\
$(0,1,0)$ &  4.695 &  1.747 &  4.157 &  -0.986 \\
$(0,0,1)$ &  4.805 &  9.437 & 26.209 &  17.941 \\
$(1,1,0)$ &  9.505 &  1.933 &  5.067 &  -5.083 \\
$(1,1,1)$ & 14.221 &  4.310 & 12.025 &  -3.825 \\
\bottomrule
\end{tabular}
\caption{Portfolio performance under fixed controls for different new business
compositions $\alpha_{\mathrm{NB}}$. The new business allocation is kept constant over time.}
\label{tab:nbonly_comparison}
\end{table}

\paragraph{Bounds definitions.}

The exact numerical values of the admissibility bounds used in each scenario are
reported in Tables~\ref{tab:bounds_static} and~\ref{tab:bounds_time}.

\begin{table}[!h]
\centering
\small
\setlength{\tabcolsep}{6pt}
\begin{tabular}{lccc}
\toprule
Scenario & $\alpha_M$ & $\alpha_L$ & $\alpha_C$ \\
\midrule
Long-only (LO) 
& $[0,\infty)$ & $[0,\infty)$ & $[0,\infty)$ \\

Constant bounds (CSTB)
& $[0.6,\,30.0]$ & $[0.9,\,10.0]$ & $[0.6,\,5.0]$ \\
\bottomrule
\end{tabular}
\caption{Time-homogeneous admissibility constraints for the portfolio weights
$(\alpha_M,\alpha_L,\alpha_C)$.}
\label{tab:bounds_static}
\end{table}

\begin{table}[!h]
\centering
\small
\setlength{\tabcolsep}{5pt}
\begin{tabular}{c|ccc|ccc}
\toprule
 & \multicolumn{3}{c|}{Lower bounds} & \multicolumn{3}{c}{Upper bounds} \\
Time $s$ 
& $\alpha_M$ & $\alpha_L$ & $\alpha_C$
& $\alpha_M$ & $\alpha_L$ & $\alpha_C$ \\
\midrule
0 & 0.8 & 0.8 & 0.6 & 2.0 & 2.0 & 2.0 \\
1 & 0.6 & 0.6 & 0.4 & 4.0 & 4.0 & 4.0 \\
2 & 0.4 & 0.4 & 0.2 & 6.0 & 6.0 & 6.0 \\
3 & 0.2 & 0.2 & 0.0 & 8.0 & 8.0 & 8.0 \\
4 & 0.0 & 0.0 & 0.0 & 10.0 & 10.0 & 10.0 \\
\bottomrule
\end{tabular}
\caption{Deterministic time-dependent admissibility constraints for the portfolio
weights $(\alpha_M,\alpha_L,\alpha_C)$ under the TDB constraint scenario.}
\label{tab:bounds_time}
\end{table}
\paragraph{Choice of bounds.}

The bound configurations considered are designed to reflect economically relevant constraint structures and to assess their impact on dynamic allocation under a finite-horizon mean--DCVaR objective.

The \emph{long-only} (LO) configuration serves as a reference case with minimal constraints. While it provides an upper benchmark in terms of performance, it also highlights a structural limitation of terminal risk constraints: since the DCVaR constraint is enforced only at maturity, the optimizer tends to concentrate risk in the final period. This leads to pronounced intertemporal risk concentration, especially in life-related components.

The \emph{constant static bounds} (CSTB) configuration isolates the effect of homogeneous exposure limits. By imposing time-invariant bounds calibrated from the long-only solution, it restricts extreme terminal allocations without fixing a temporal pattern. As a result, the optimizer is forced to engage earlier in the horizon, leading to a more balanced intertemporal allocation.

The \emph{time-dependent bounds} (TDB) configuration further refines this approach by allowing admissible exposure ranges to evolve over time. These bounds are designed to mimic progressive portfolio rebalancing and gradual repositioning of activities, as typically observed in (re)insurance practice. As in the CSTB case, this reduces terminal risk concentration, but introduces a more realistic temporal structure. In particular, early-period allocations are constrained toward lower bounds, while later-period allocations gradually approach upper limits.

Overall, these bound structures illustrate how intertemporal constraints shape optimal policies under explicit DCVaR constraints. In particular, unconstrained solutions tend to exploit finite-horizon effects, whereas structured bounds lead to more stable and economically consistent allocation profiles.

Tables~\ref{tab:grid_if00_train_summary} and~\ref{tab:grid_if11_train_summary} report results on both training and evaluation datasets. For neural network policies, constraint levels on the training set lie slightly below $K$, reflecting the boundary calibration procedure described in Subsection~\ref{subsubsec:numerical_results}.

\paragraph{Constraint Tightness and Learning Gains.}

Table~\ref{tab:alpha_fixed_bounds_both} reports the optimal fixed (time-deterministic) exposure profiles under each constraint configuration. These deterministic optima provide a direct measure of constraint tightness: when they lie on the boundary of the feasible set, the admissible region is effectively saturated and the scope for dynamic improvement is limited.

Under both TDB and CSTB specifications, this effect is particularly visible in the IF00 configuration. The fixed solution remains close to the admissible boundaries, leaving little room for intertemporal reallocation and resulting in only marginal gains from the neural policy. In contrast, under IF11, the presence of an initial in-force portfolio alters the exposure balance and prevents systematic boundary saturation. The feasible region is therefore less compressed, allowing the neural policy to exploit residual flexibility and achieve significant improvements.

In the LO configuration, the absence of upper bounds considerably enlarges the feasible set. The neural policy can fully leverage dynamic flexibility, leading to pronounced performance gains, with the impact of the in-force composition becoming secondary.

Overall, these results show that the effective restrictiveness of the constraint set depends not only on the bounds themselves but also on the initial portfolio composition. For strictly comparable experiments, admissibility bounds should therefore be calibrated conditional on the in-force portfolio in order to equalize constraint pressure across configurations.

\begin{table}[!h]
\centering
\small
\setlength{\tabcolsep}{4.5pt}
\begin{tabular}{c c c | ccc | ccc}
\hline
Config. & $t$ & $[\alpha^{\mathrm{low}}_t,\alpha^{\mathrm{up}}_t]$
& \multicolumn{3}{c|}{IF00} & \multicolumn{3}{c}{IF11} \\
\cline{4-9}
 &  &  & $\alpha^M_t$ & $\alpha^L_t$ & $\alpha^C_t$ & $\alpha^M_t$ & $\alpha^L_t$ & $\alpha^C_t$ \\
\hline
\multirow{5}{*}{LO}
 & 0 & $[0,\infty)$
 & $0.00$ & $3.63$ & $0.44$ & $0.00$ & $0.00$ & $0.52$ \\
 & 1 & $[0,\infty)$
 & $0.00$ & $0.71$ & $0.43$ & $0.00$ & $0.00$ & $0.41$ \\
 & 2 & $[0,\infty)$
 & $0.00$ & $0.26$ & $0.27$ & $0.00$ & $0.00$ & $0.32$ \\
 & 3 & $[0,\infty)$
 & $0.00$ & $1.91$ & $0.57$ & $0.00$ & $0.00$ & $0.37$ \\
 & 4 & $[0,\infty)$
 & $37.98$ & $31.20$ & $0.59$ & $2.94$ & $22.86$ & $0.32$ \\
\hline
\multirow{5}{*}{CSTB}
 & 0 & $[0.6,30]\times[0.9,10]\times[0.6,10]$
 & $0.60$ & $4.21$ & $0.71$ & $0.60$ & $0.90$ & $0.60$ \\
 & 1 & $[0.6,30]\times[0.9,10]\times[0.6,10]$
 & $0.60$ & $1.09$ & $0.60$ & $0.60$ & $0.90$ & $0.60$ \\
 & 2 & $[0.6,30]\times[0.9,10]\times[0.6,10]$
 & $0.60$ & $4.47$ & $0.60$ & $0.60$ & $0.90$ & $0.60$ \\
 & 3 & $[0.6,30]\times[0.9,10]\times[0.6,10]$
 & $6.20$ & $10.00$ & $0.75$ & $0.60$ & $4.25$ & $0.60$ \\
 & 4 & $[0.6,30]\times[0.9,10]\times[0.6,10]$
 & $30.00$ & $10.00$ & $0.68$ & $0.60$ & $10.00$ & $0.60$ \\
\hline
\multirow{5}{*}{TDB}
 & 0 & $[0.8,2]\times[0.8,2]\times[0.6,2]$
 & $2.00$ & $2.00$ & $1.11$ & $0.80$ & $0.80$ & $0.60$ \\
 & 1 & $[0.6,4]\times[0.6,4]\times[0.4,4]$
 & $4.00$ & $4.00$ & $0.95$ & $0.60$ & $0.60$ & $0.55$ \\
 & 2 & $[0.4,6]\times[0.4,6]\times[0.2,6]$
 & $6.00$ & $6.00$ & $0.87$ & $0.40$ & $0.40$ & $0.24$ \\
 & 3 & $[0.2,8]\times[0.2,8]\times[0,8]$
 & $8.00$ & $8.00$ & $1.14$ & $0.20$ & $5.70$ & $0.47$ \\
 & 4 & $[0,10]\times[0,10]\times[0,10]$
 & $10.00$ & $10.00$ & $1.02$ & $1.77$ & $10.00$ & $0.31$ \\
\hline
\end{tabular}
\caption{Optimal fixed exposure weights $\alpha^{\mathrm{fixed}}$ under different bound
configurations. Results are reported for $\alpha_{\mathrm{IF}}=(0,0,0)$ (IF00) and
$\alpha_{\mathrm{IF}}=(1,1,0)$ (IF11). Bounds are reported per $(t,\mathrm{LoB})$ and are common to IF00 and IF11.
Values are reported with two decimals.}
\label{tab:alpha_fixed_bounds_both}
\end{table}

\begin{table}[!h]
\centering
\small
\setlength{\tabcolsep}{6pt}
\begin{tabular}{lllrrrr}
\toprule
Bounds & Policy & Set &
$\mathbb{E}[R]$ & $\mathrm{Std}(R)$ &
$\mathrm{DCVaR}_{0.99}(-R)$ & $\mathrm{VaR}_{0.99}(-R)$ \\
\midrule
\multirow{6}{*}{LO}
 & NN    & Train & \textbf{97.721} & 16.967 & 29.269 & -72.783 \\
 & NN    & Eval  & \textbf{98.280} & 17.718 & 30.030 & -72.799 \\
 & FT    & Train & 71.684 & 11.338 & 30.037 & -45.514 \\
 & FT    & Eval  & 71.671 & 11.385 & 30.128 & -45.425 \\
 & Const & Train & 57.433 & 11.304 & 29.946 & -31.336 \\
 & Const & Eval  & 57.422 & 11.380 & 30.150 & -31.152 \\
\midrule
\multirow{6}{*}{CSTB}
 & NN    & Train & \textbf{81.865} & 11.874 & 29.933 & -56.395 \\
 & NN    & Eval  & \textbf{81.861} & 11.912 & 30.080 & -56.306 \\
 & FT    & Train & 66.076 & 11.200 & 30.001 & -40.001 \\
 & FT    & Eval  & 66.062 & 11.291 & 30.224 & -39.746 \\
 & Const & Train & 57.433 & 11.303 & 29.946 & -31.337 \\
 & Const & Eval  & 57.422 & 11.379 & 30.150 & -31.150 \\
\midrule
\multirow{6}{*}{TDB}
 & NN    & Train & \textbf{60.326} & 11.140 & 29.547 & -34.614 \\
 & NN    & Eval  & \textbf{60.312} & 11.276 & 29.864 & -34.265 \\
 & FT    & Train & 60.329 & 11.275 & 29.996 & -34.238 \\
 & FT    & Eval  & 60.344 & 11.286 & 29.898 & -34.254 \\
 & Const & Train & 28.376 & 10.183 & 29.057 &  -3.456 \\
 & Const & Eval  & 28.433 & 10.196 & 28.851 &  -3.544 \\
\bottomrule
\end{tabular}
\caption{Training-set performance for Inforce configuration
$\alpha_{IF}=(0,0,0)$ under different constraint sets.
Reported are the mean and standard deviation of the reward, the empirical
$\mathrm{DCVaR}_{0.99}$, and the $\mathrm{VaR}_{0.99}$ of the loss.}
\label{tab:grid_if00_train_summary}
\end{table}

\begin{table}[!h]
\centering
\small
\setlength{\tabcolsep}{6pt}
\begin{tabular}{lllrrrr}
\toprule
Bounds & Policy & Set &
$\mathbb{E}[R]$ & $\mathrm{Std}(R)$ &
$\mathrm{DCVaR}_{0.99}(-R)$ & $\mathrm{VaR}_{0.99}(-R)$ \\
\midrule
\multirow{6}{*}{LO}
 & NN    & Train & \textbf{91.078} & 23.697 & 28.651 & -66.646 \\
 & NN    & Eval  & \textbf{92.078} & 25.529 & 30.031 & -66.661 \\
 & FT    & Train & 24.317 & 11.085 & 29.996 &   1.804 \\
 & FT    & Eval  & 24.327 & 11.095 & 29.770 &   1.651 \\
 & Const & Train & 16.680 & 11.157 & 29.996 &   9.389 \\
 & Const & Eval  & 16.703 & 11.173 & 29.871 &   9.336 \\
\midrule
\multirow{6}{*}{CSTB}
 & NN    & Train & \textbf{53.660} & 12.985 & 29.594 & -29.661 \\
 & NN    & Eval  & \textbf{53.707} & 13.191 & 30.006 & -29.473 \\
 & FT    & Train & 20.117 & 10.990 & 29.996 &   5.969 \\
 & FT    & Eval  & 20.140 & 11.003 & 29.829 &   5.856 \\
 & Const & Train & 16.287 & 11.014 & 29.996 &   9.699 \\
 & Const & Eval  & 16.316 & 11.033 & 29.916 &   9.733 \\
\midrule
\multirow{6}{*}{TDB}
 & NN    & Train & \textbf{41.257} &  9.725 & 29.736 & -17.418 \\
 & NN    & Eval  & \textbf{41.242} &  9.776 & 30.198 & -17.098 \\
 & FT    & Train & 20.706 & 10.996 & 29.997 &   5.354 \\
 & FT    & Eval  & 20.724 & 11.008 & 29.809 &   5.234 \\
 & Const & Train & 15.927 & 10.944 & 29.996 &  10.096 \\
 & Const & Eval  & 15.957 & 10.963 & 29.935 &  10.092 \\
\bottomrule
\end{tabular}
\caption{Training-set performance for Inforce configuration
$\alpha_{IF}=(1,1,0)$ under different constraint sets.
Reported are the mean and standard deviation of the reward, the empirical
$\mathrm{DCVaR}_{0.99}$, and the $\mathrm{VaR}_{0.99}$ of the loss.}
\label{tab:grid_if11_train_summary}
\end{table}

\printbibliography

\end{document}